\documentclass[twocolumn]{svjour3}  
\smartqed  

\usepackage[dvipdfmx]{color}
\usepackage{graphicx}
\usepackage[T1]{fontenc}
\usepackage{mathptmx}  
\usepackage[scaled]{helvet} 
\usepackage{courier}
\usepackage[subrefformat=parens]{subcaption}

\makeatletter
\let\cl@chapter\undefined
\makeatletter
\usepackage{amsmath,amssymb}   
\usepackage{mathtools} 
\usepackage[colorlinks=true,linkcolor=cyan,urlcolor=blue,citecolor=cyan]{hyperref}
\usepackage{cleveref}  
\crefname{equation}{Eq.}{Eqs.}%
\crefname{figure}{Fig.}{Figs.}%
\usepackage{amsbsy}
\usepackage{bm}
\usepackage{algorithmic}
\usepackage{algorithm}
\usepackage{enumerate}
\usepackage{ragged2e}
\usepackage{caption,subcaption }
\usepackage[toc,page]{appendix}
\usepackage{booktabs}
\usepackage{multirow}

\usepackage{xcolor}
\usepackage{pgfplots}
\usepackage{cite}

\newcommand{\uvelo}{u_{\mathrm{s}}}
\newcommand{\uveloO}{u_{\mathrm{s}0}}
\newcommand{\uvelodot}{\dot{u}_{\mathrm{s}} }
\newcommand{\uvelodotO}{\dot{u}_{\mathrm{s}0} }
\newcommand{\vm}{v_{\mathrm{m}}}
\newcommand{\vmO}{v_{\mathrm{m}0}}
\newcommand{\vmdot}{\dot{v}_{\mathrm{m}}}
\newcommand{\vmdotO}{\dot{v}_{\mathrm{m}0}}
\newcommand{\rdot}{\dot{r}}
\newcommand{\rdotO}{\dot{r}_{0}}

\newcommand{\xG}{x_{\mathrm{G}}}
\newcommand{\momentN}{N_{\mathrm{m}}}

\newcommand{\UT}{U_{\mathrm{T}}}
\newcommand{\UA}{U_{\mathrm{A}}}
\newcommand{\gammaT}{\gamma_{\mathrm{T}}}
\newcommand{\gammaA}{\gamma_{\mathrm{A}}}
\newcommand{\deltas}{\delta_{\mathrm{s}}}
\newcommand{\deltap}{\delta_{\mathrm{p}}}
\newcommand{\deltaptilde}{\tilde{\delta}_{\mathrm{p}}}
\newcommand{\deltastilde}{\tilde{\delta}_{\mathrm{s}}}
\newcommand{\ntilde}{\tilde{n}}
\newcommand{\iBTtilde}{\tilde{I}_{\mathrm{BT}}}
\newcommand{\deltaphover}{\delta_{\mathrm{p, hover}}}
\newcommand{\deltashover}{\delta_{\mathrm{s, hover}}}

\newcommand{\degree}{^{\circ}}
\newcommand{\Ibt}{I_\mathrm{BT}}
\newcommand{\Ibto}{I_\mathrm{BT0}}

\newcommand{\tfpla}{t_{\mathrm{f}}}
\newcommand{\lpp}{L_\mathrm{pp}}

\newcommand{\dt}{\mathrm{d}t}

\newcommand{\qsimI}{q^{i}_{\mathrm{sim}}}
\newcommand{\qhatsimI}{\hat{q}^{i}_{\mathrm{sim}}}

\newcommand{\qinI}{q^{i}_{\mathrm{in}}}
\newcommand{\qhatinI}{\hat{q}^{i}_{\mathrm{in}}}
\newcommand{\cospsi}{\cos{\psi}}
\newcommand{\sinpsi}{\sin{\psi}}
\newcommand{\sumstate}{\sum_{i=1}^{N}}
\newcommand{\norm}[1]{\lVert#1\rVert}   
\newcommand{\lpare}[1]{\left(#1\right)} 

\crefname{table}{Table}{Tables}%
\crefname{equation}{Eq.}{Eq.}%
\crefname{figure}{Fig.}{Fig.}%
\crefname{algorithm}{Algorithm}{Algorithm}%
\crefname{section}{Section}{Section}
\creflabelformat{section}{#2#1#3}
\crefname{subsection}{Section}{Section}
\creflabelformat{subsection}{#2#1#3}
\crefname{subsubsection}{Section}{Section}
\creflabelformat{subsubsection}{#2#1#3}

\usepackage{ulem}

\providecommand{\Erase}{\bgroup\markoverwith{\textcolor{red}{\rule[.5ex]{2pt}{0.4pt}}}\ULon}

\usepackage{etoolbox}
\newcommand{\affaddr}[1]{#1} 
\newcommand{\affmark}[1][*]{\textsuperscript{#1}}
\usepackage[misc]{ifsym}

\journalname{Journal of Marine Science and Technology}
\begin{document}

\title{ Development and Identification of a Linear Low-Speed Ship Maneuvering Model from Full-Scale Data

}

\author{
Agnes N. Mwange\protect\affmark[1,2*] \and 
Taichi Kambara\protect\affmark[1] \and 
Kouki Wakita\protect\affmark[1] \and 
Kazuyoshi Hosogaya\protect\affmark[3] \and
Atsuo Maki\protect\affmark[1*]
}

\authorrunning{
Agnes N. Mwange \and 
Taichi Kambara \and 
Kazuyoshi Hosogaya\and
Atsuo Maki \and 
}

\institute{
            \affaddr{\affmark[1]{Department of Naval Architecture and Ocean Engineering, Graduate School of Engineering, Osaka University, Suita, Osaka, Japan. }} 
            \\\\
            \affaddr{\affmark[2]{Department of Marine Engineering and Maritime Operations, Jomo Kenyatta University of Agriculture and Technology (JKUAT), Kenya.}}\\\\  
            \affaddr{\affmark[3]{Japan Hamworthy Co., Ltd.}}\\\\ 
            \affmark[*]{Corresponding authors}\\
            \Letter $~$ Agnes N. Mwange\\ $~~~~~~~${mwange\_agnes\_ngina@naoe.eng.osaka-u.ac.jp \\\
            $~~~~~~~$Atsuo Maki\\ 
            $~~~~~~~$maki@naoe.eng.osaka-u.ac.jp}
}

\date{Received: date / Accepted: date}

\maketitle
\begin{abstract}
Despite significant technological progress, the realization of fully autonomous berthing and unberthing remains a significant challenge. One of the primary obstacles is the complex, non-linear nature of low-speed ship dynamics, which are difficult to model and control and often necessitate equally complex maneuvering models and control systems. This study proposes a simplified approach to bridge this gap by modeling the ship dynamics in the form of a time-invariant, continuous-time linear state-space system. The model parameters are estimated through system identification using the Covariance Adaptation Strategy Evolution Strategy (CMA-ES) applied to full-scale maneuvering data. Validation results demonstrate a strong agreement between the model output and empirical data. This outcome demonstrates the significant potential of simplified models to effectively define the maneuvering motion of a ship at low speeds.

\end{abstract}
\section{Introduction}\label{sec: introduction}
Mathematical models for low-speed ship maneuvering are predominantly classified into two: polynomial models and hydrodynamic models. Polynomial models \cite{Abkowitz1964a, Strom-Tejsen1965a, measurement_abkowitz_1980} treat the ship as a rigid body. The equations of motion are derived via Taylor series expansion of the state variables and control inputs, where the model's complexity and accuracy are governed by the order of this expansion. The primary advantage of this approach is its parametric simplicity, as it does not require explicit consideration of hydrodynamic interactions between ship components. In contrast, hydrodynamic models such as the Maneuvering Model Group (MMG) models \cite{Ogawa1978b, Yoshimura1978, hydrodynamic_inoue_1981, mathematical_kose_1984, unified_yoshimura_2009, introduction_yasukawa_2015}, adopt a modular approach by synthesizing the contributions of the ship's principal components such as hull, propeller and rudder. The equations of motion are formulated by integrating the forces and moments of each component, providing clear physical transparency into their interactions. However, this physical detail necessitates the identification of a large parameter set and involves complex modeling of component interactions. Generally, the selection of the appropriate approach for a given application involves a trade-off between computational efficiency, desired accuracy, and the required level of physical interpretability.
 
\subsection{Related Research}\label{sec: related research}
The development of reliable ship maneuvering models for automation has historically relied on empirical data, facing a persistent trade-off between complexity, interpretability, and accurate model-to-full-scale extrapolation.

 Early foundational work established the Taylor-series expansion model by Abkowitz et al. \cite{Abkowitz1964a} and parameter identification via captive model tests (CMTs) using planar motion mechanisms \cite{Strom-Tejsen1965a}. Subsequent studies on specialized maneuvers based on model tests revealed limitations in simulating full-scale propulsion dynamics \cite{Yoshimura1978,FUJINO1978}. The modular MMG framework \cite{Ogawa1978b, Kose1979a} improved physical interpretability and provided a structured methodology that enabled the determination of hydrodynamic parameters for principal ship components (hull, rudder and propeller) through CMTs. The robustness of this approach was proven through significant extensions to mathematical models for shallow water incorporating sinkage and trim effects \cite{YOSHIMURA1988, YOSHIMURA1989}, high-speed to low-speed maneuvering \cite{unified_yoshimura_2009}, and formulations for twin-propeller twin-rudder ships \cite{ Lee1988, KOBAYASHI1994, aspects_dubbioso_2012}. A principal limitation of this approach is the difficulty in directly measuring interaction coefficients between components, necessitating empirical estimation that may itself be subject to scale effects. Further on, full-scale trials with the Esso Osaka \cite{Crane1979} revealed deficiencies in existing model identification procedures and underscored the significance of scale effects, particularly in shallow water conditions. Abkowitz et al. \cite{Abkowitz1980} mitigated uncertain scaling effects by deriving parameters directly from full-scale trials; however, this method is constrained by the high cost and operational complexity of conducting such trials for every ship. Alternative approaches prioritized practical applications: Hirano et al. \cite{HIRANO1985} emphasized practical calculation methods for initial design, whereas Biancardi et al. \cite{simplified_biancardi_1988} prioritized computational efficiency in developing simplified models for onboard simulators.

To address inherent scale effects, subsequent research by Ueno et al. \cite{Ueno2001, Ueno2014a, Ueno2015a, Ueno2017} employed free-running model tests with auxiliary thrusters, applying corrections to rudder effectiveness and speed response to achieve full-scale-equivalent motion. This approach further demonstrated the increased complexity required for model tests to accurately represent full-scale dynamics. Further studies on the MMG model \cite{introduction_yasukawa_2015} utilizing CMTs acknowledged the continued reliance on empirical regression formulas for model-to-full-scale translation.

Alternative approaches have sought to circumvent traditional model tests. One approach utilizes Computational Fluid Dynamics (CFD) to perform "virtual" CMTs \cite{Liu2018, Sakamoto2019}. Primary constraints of this method include significant computational cost, uncertainties in turbulence modeling, and the challenge of simulating rotating propellers and moving rudders. A second approach employs time-domain System Identification (SI), pioneered by Abkowitz et al. \cite{measurement_abkowitz_1980} and Hwang et al. \cite{ Hwang1980, Hwang1982}, which treats parameter estimation as an optimization problem—minimizing error between simulated and recorded ship trajectories from standard maneuvers (e.g., zigzag tests). Modern machine learning techniques extend this core SI principle, utilizing operational data to identify parameters for parametric models \cite{Miyauchi2022SI, data_deogaonkar_2023} or to learn entirely data-driven models \cite{ discovering_hasan_2025}. For instance, identification from random maneuver data has demonstrated superior agreement with experimental results compared to traditional CMT-derived models \cite{Miyauchi2022SI}. This work also highlighted that models identified from standard maneuvers, such as turning circles and zigzags, may not generalize well to low-speed berthing maneuvers. Ultimately, the efficacy of data-driven parameter identification is highly dependent on the quality and quantity of the training data.

In summary, a significant research gap persists in the formulation of reliable, physically simplified maneuvering models capable of accurately capturing full-scale performance without incurring high computational cost.

\subsection{Research Objectives and Overview}
This study aims to develop a simplified low-speed maneuvering model that does not compromise on accuracy and full-scale interpretability. The model parameters are identified directly from full-scale operational data using CMA-ES. While CMA-ES is often associated with high computational cost, the reduced parameter set of the proposed model renders it an exceptional tool for obtaining optimal parameters with significantly lower computational cost. As detailed in \cref{sec: data curation}, the training dataset is curated to mitigate the parameter cancellation effect, a phenomenon in system identification where strong correlations between state variables, such as yaw rate ($r$) and sway velocity ($v_\mathrm{m}$), can lead to numerically indeterminate and physically inconsistent estimates of hydrodynamic coefficients \cite{Hwang1982, measures_luo_2017}.

\subsection{Notations}\label{sec: notaions}
This section provides definitions of the symbols used throughout this study. The $n$-dimensional Euclidean space is denoted by $\mathbb{R}^n$ while the set of real numbers for  $n=1$ is represented by $\mathbb{R}$.  

\section{Methods} \label{sec: methods}
\subsection{Subject Ship}
\begin{figure}[htbp]
    \centering
    \includegraphics[width = 1.0\columnwidth]{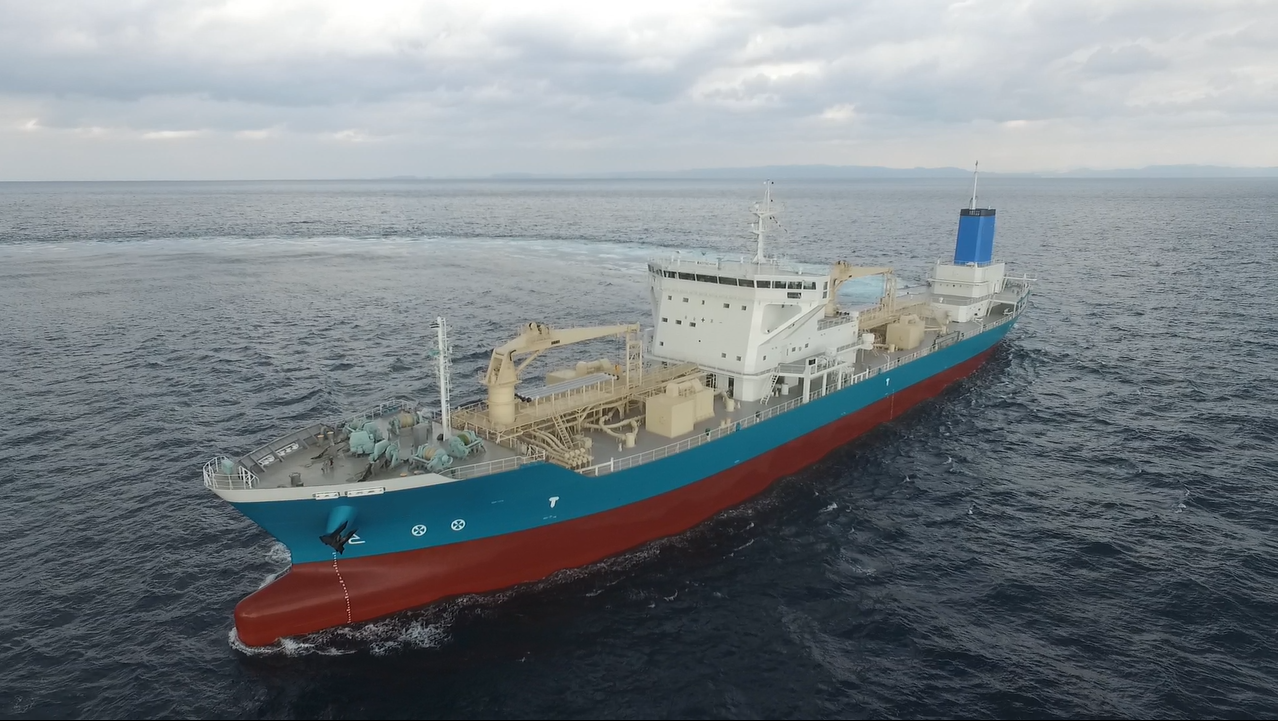}
    \caption{Subject ship used in the study \cite{Mwange2025}.}
    \label{fig: full-scale ship}
\end{figure}

The subject ship, shown in \cref{fig: full-scale ship}, is a coastal ship equipped with a vectwin rudder system and a controllable pitch bow thruster. 

The principal particulars of the ship are detailed in \cref{tab: principal particulars}.

\begin{table}[htbp]
    \setlength{\tabcolsep}{4.5pt}
    \captionsetup{skip=0pt,singlelinecheck=off, justification=raggedright}
    \caption{Principal particulars of the subject ship.  }
    \begin{tabular}{p{30mm} p{40mm}}
    \toprule
      Parameter   & Detail \\
    \hline\hline  
      Length ($\lpp$)   & Approx. 150 m \\
      Breadth ($B$)  & Approx. 25m \\
      Draft ($d$) &  Approx. 8.6m \\
      Rudder & Vectwin rudder system\\
      Propeller & 1 fixed pitch propeller (FPP)\\
      Side thrusters & 1 controllable pitch bow thruster\\
      \bottomrule
    \end{tabular}
    \label{tab: principal particulars}
\end{table}

\subsection{Maneuvering Model}
\begin{figure}[htbp]
    \centering
    \includegraphics[width = 1.08\columnwidth]{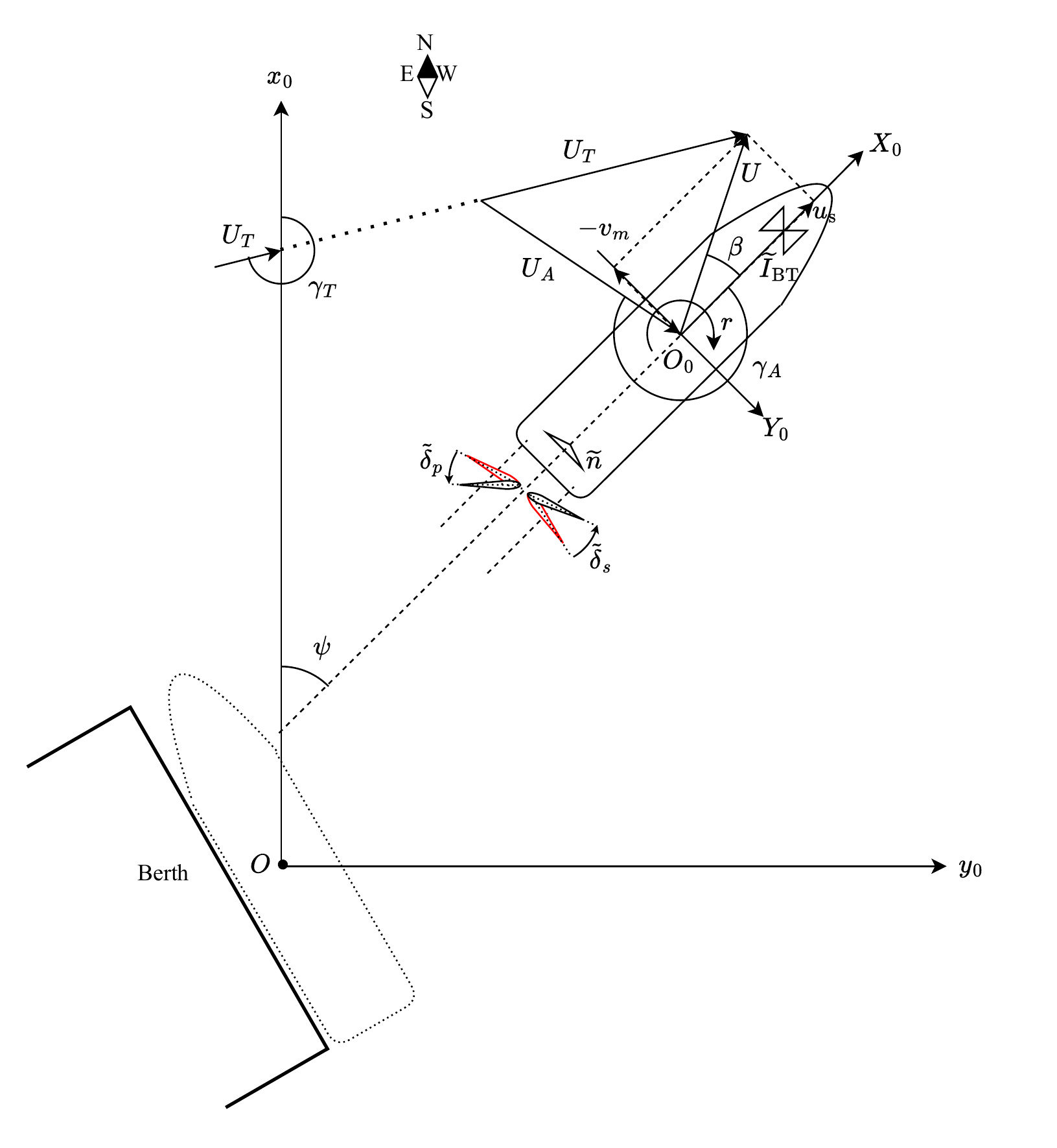}
    \caption{Coordinate systems.}
    \label{fig: full scale coordinate systems}
\end{figure}

This study employs a 3-degree-of-freedom (3DOF) kinematic model, defined within two principal coordinate systems: an inertial frame (earth-fixed coordinate system), denoted as  $O-x_{0}y_{0}$ and a ship-fixed coordinate system, denoted as $O_{0}-X_{0}Y_{0}$, as illustrated in \cref{fig: full scale coordinate systems}. The origin of the ship-fixed coordinate system, $O_{0}$, is set at the ship's center of gravity. The origin of the inertial coordinate system, $O$, coincides with $O_{0}$ when the ship is at the berth. 

The relationship between the two coordinate systems is governed by:
\begin{equation}
  \begin{bmatrix} \dot{x}_0 \\ \dot{y}_0 \\ \dot{\psi} \end{bmatrix} =
  \begin{bmatrix} \cospsi & -\sinpsi &  0 \\ \sinpsi & \cospsi & 0 \\  0 & 0 & 1 \end{bmatrix} \begin{bmatrix} \uvelo\\ \vm \\ r  \end{bmatrix}  
  \label{eq: earth-ship coordinate systems}
\end{equation}
where $ \dot{x}_{0}, \dot{y}_0, \dot{\psi} $ denote the time derivative of the ship's position in the x-axis and y-axis, respectively, and $\psi$ denotes the ship's heading angle in the inertial frame. The terms $\uvelo, \vm$ and $r$ denote the ship's surge, sway and yaw velocities defined in the ship-fixed coordinate system, respectively.

The ship is equipped with a vectwin rudder system, enabling the ship to achieve a hover state/mode (stationary under constant propeller thrust) through specific port ($\delta_p$) and starboard ($\delta_s$) rudder angle combinations, (typically $\delta_p = -75^\circ$, $\delta_s = 75^\circ$). In this state, the ship can perform linearized motions, including crabbing (pure lateral translation) \cite{rachman2023experimental}. Consequently, this study defines the hover condition as the initial equilibrium point for deriving the linearized ship dynamics. When hovering, the ship's center of gravity is located at midships, such that the distance of the ship's center of gravity from midships, $x_G$, is zero.

Generally, the nonlinear equations of motion of the ship about the ship's center of gravity, $\xG$, in the ship-fixed coordinate system are defined as follows:
\begin{equation}
       \begin{split}
       (m + m_x)\uvelodot - (m + m_y) \vm r - \xG m r ^ 2 &= X   \\       
       (m + m_y)\vmdot+ (m + m_x) \uvelo r + \xG m \dot{r} &= Y \\        
       \lpare{I_{zz} + J_{zz} + \xG m ^ 2}\dot{r} + \lpare{\vm + \uvelo r}\xG m &= \momentN  \\  \end{split}
    \label{eq: equations of motion}
\end{equation}
where $m$ refers to the ship's mass while  $m_x, m_y $ denote the added mass coefficients in the 
x- and y-axes, respectively. Similarly, $I_{zz}$ and $J_{zz}$ represent the ship's moment of inertia and added moment of inertia, both referenced about $\xG$. On the right-hand side of \cref{eq: equations of motion}, $X$ and $Y$ denote the total surge and sway forces, respectively, whereas $\momentN$  denotes the total yaw moment about the midships.

By neglecting the nonlinear and zero-valued terms in \cref{eq: equations of motion} such as $\vm r, \; \uvelo r$ , the linearized equations of motion are derived as follows:
\begin{equation}
       \begin{split}
       (m + m_x)\uvelodot &= X   \\       
       (m + m_y)\vmdot &= Y \\        
       \lpare{I_{zz} + J_{zz}} \dot{r} &= \momentN  \\  \end{split}
    \label{eq: linear equations of motion}
\end{equation}
Further, the right-hand-side of \cref{eq: linear equations of motion} can be decomposed as:

\begin{equation} \label{eq: water and air linear equations of motion}
       \begin{split}
       X&= X_{\mathrm{water}} +  X_{\mathrm{air}}  \\       
       Y &= Y_{\mathrm{water}} +  Y_{\mathrm{air}}  \\        
       \momentN &= N_{\mathrm{water}} +  N_{\mathrm{air}}  \\  
       \end{split}
\end{equation}
where the subscript 'water' denotes hydrodynamic forces and moments, while the subscript 'air' denotes wind-induced forces and moments. The hydrodynamic forces and moment are functions of the ship's motion parameters and control inputs \cite{Abkowitz1964a,Strom-Tejsen1965a} as expressed below:
\begin{equation} \label{eqn: forces and moments as func. of }
    \begin{split}
     X_{\mathrm{water}} = X(\uvelo, \vm, r, \uvelodot, \vmdot, \rdot,  \deltaptilde, \deltastilde, \ntilde, \iBTtilde)\\
     Y_{\mathrm{water}} = Y(\uvelo, \vm, r, \uvelodot, \vmdot, \rdot,  \deltaptilde, \deltastilde, \ntilde, \iBTtilde)\\
    N_{\mathrm{water}} = \momentN (\uvelo, \vm, r, \uvelodot, \vmdot, \rdot,  \deltaptilde, \deltastilde, \ntilde, \iBTtilde)
    \end{split}
\end{equation}
where $\deltaptilde, \deltastilde, \ntilde, \iBTtilde$ denote the deviations of the port rudder angle, starboard rudder angle, propeller revolutions, and bow thruster current, respectively, from the initial equilibrium condition, defined as follows:

\begin{equation} \label{eqn: control dif. hover}
    \begin{split}
     \deltaptilde &\equiv \deltap - \deltaphover \\
     \deltastilde & \equiv \deltas - \deltashover \\
     \ntilde & \equiv n - n_{0}\\
     \iBTtilde &\equiv (\Ibt - \Ibto)/ I_{\mathrm{BT, amp}}    
    \end{split}
\end{equation}
where the absolute port and starboard rudder hover angles range between $70^\circ$ and $80^\circ$, that is, $|\deltaphover|, |\deltashover| \in [70^\circ, 80^\circ]$. $n_0 = 1.2 \; [\mathrm{rps}]$, $\Ibto = 12.0 \; [\mathrm{mA}]$ and $I_{\mathrm{BT, amp}} = 8.0 \; [\mathrm{mA}]$.

Now, starting with $X_\mathrm{water}$, considering the change in any state variable is defined in the form $\triangle{x} = x - x_{0}$, and the change in the control inputs is defined according to \cref{eqn: control dif. hover}, a first-order Taylor series expansion of the force at any instant about the initial equilibrium condition yields: 
\begin{equation}\label{eqn: X taylor expansion}
\begin{split}
    X_{\mathrm{water}} &= X_{0} + \frac{\partial{X}}{\partial{\uvelo}}\triangle{\uvelo} + \frac{\partial{X}}{\partial{\vm}}\triangle{\vm} + \frac{\partial{X}}{\partial{r}}\triangle{r}  \\ &+  \frac{\partial{X}}{\partial{\uvelodot}}\triangle{\uvelodot} +  \frac{\partial{X}}{\partial{\vmdot}}\triangle{\vmdot} + \frac{\partial{X}}{\partial{\rdot}}\triangle{\rdot}
    \\ &+  \frac{\partial{X}}{\partial{\deltaptilde}}\deltaptilde + \frac{\partial{X}}{\partial{\deltastilde}}\deltastilde + \frac{\partial{X}}{\partial{\ntilde}}\ntilde + \frac{\partial{X}}{\partial{\iBTtilde}}\iBTtilde
\end{split}
\end{equation}

Moreover, the ship is stationary in the hover position; therefore, the equilibrium states, $\uveloO, \vmO, r_0$ and state derivatives, $\uvelodotO, \vmdotO, \rdotO $ are zero. Consequently, the change in any state variable simplifies to $\triangle{x}= x$. Additionally, the hydrodynamic forces due to control inputs are balanced at equilibrium, resulting in a stationary condition, which implies:  $X_0  = X(0,0,\dots, 0) = 0$. By denoting the partial derivatives as $\frac{\partial{X}}{\partial{i}} \equiv X_i $, the Taylor series expansion for 
 $X_{\mathrm{water}}$ simplifies to:
\begin{equation}
    \begin{split}
       X_{\mathrm{water}} =\; & X_{\uvelo}\uvelo + X_{\vm}\vm + X_{r}r + X_{\uvelodot}\uvelodot + X_{\vmdot}\vmdot +X_{\rdot}\rdot  \\&+  X_{\deltaptilde}\deltaptilde + X_{\deltastilde}\deltastilde + X_{\ntilde}\ntilde + X_{\iBTtilde}\iBTtilde
    \end{split}  
\end{equation}

For a ship with a symmetric hull, the following hydrodynamic derivatives vanish due to symmetry \cite{Strom-Tejsen1965a}:
\begin{equation}
  \{ X_{\vm}, X_{r},  X_{\vmdot}, X_{\rdot}\} = 0,   
\end{equation}
and the bow thruster has no direct effect on surge, so 
 $X_{\iBTtilde} = 0$. Similarly, the expansions for $Y_{\mathrm{water}}, N_{\mathrm{water}}$ are derived in an analogous manner and with the symmetry conditions:
 \begin{equation}
     \{ Y_{\uvelo}, Y_{\uvelodot}, N_{\uvelo}, N_{\uvelodot} \} = 0. 
 \end{equation}
 
 Consequently, the final first-order expansions of the hydrodynamic forces and moment are given by:
\begin{equation} \label{eqn: final taylor expansion}
    \begin{split}
       X_{\mathrm{water}} &= X_{\uvelo}\uvelo + X_{\uvelodot}\uvelodot + X_{\deltaptilde}\deltaptilde + X_{\deltastilde}\deltastilde + X_{\ntilde}\ntilde \\
       Y_{\mathrm{water}} &= Y_{\vm}\vm + Y_{r}r  + Y_{\vmdot}\vmdot + Y_{\rdot}\rdot  \\&+  Y_{\deltaptilde}\deltaptilde + Y_{\deltastilde}\deltastilde + Y_{\ntilde}\ntilde + Y_{\iBTtilde}\iBTtilde\\
      N_{\mathrm{water}} &= N_{\vm}\vm + N_{r}r  + N_{\vmdot}\vmdot + N_{\rdot}\rdot  \\&+  N_{\deltaptilde}\deltaptilde + N_{\deltastilde}\deltastilde + N_{\ntilde}\ntilde + N_{\iBTtilde}\iBTtilde    
    \end{split}  
\end{equation}

Wind-induced forces and moment in \cref{eq: water and air linear equations of motion} are defined using Fujiwara's regression formulas as follows:
\begin{equation} \label{eqn: air force and moments}
    \begin{split}
       X_{\mathrm{air}} &= \frac{1}{2}\rho_{\mathrm{A}}\UA^{2}A_{\mathrm{T}}.C_{X}\\
       Y_{\mathrm{air}} &= \frac{1}{2}\rho_{\mathrm{A}}\UA^{2}A_{\mathrm{L}}.C_{Y}\\
       N_{\mathrm{air}} &= \frac{1}{2}\rho_{\mathrm{A}}\UA^{2}A_{\mathrm{L}}L_{\mathrm{OA}}.C_{N}
    \end{split}  
\end{equation}
where $\rho_{\mathrm{A}}$ is the air density, $\UA$ is the relative wind speed, $A_{\mathrm{T}}$ is the transverse-projected windage area, $A_{\mathrm{L}}$ is the longitudinal-projected windage area, and $L_{\mathrm{OA}}$ is the overall ship length. The coefficients $C_{X}, C_{Y}, C_{N}$ are functions of the relative wind direction $\gammaA$ and defined as follows:
\begin{equation}\label{eqn: wind coeffs}
    \begin{split}
       C_{X} = &X_{\mathrm{A0}} + X_{\mathrm{A1}} \cos\lpare{{2\pi - \gammaA}} + X_{\mathrm{A3}} \cos3\lpare{{2\pi - \gammaA}} \\&+ X_{\mathrm{A5}} \cos5\lpare{{2\pi - \gammaA}} \\
       C_{Y} = &Y_{\mathrm{A1}} \cos\lpare{{2\pi - \gammaA}} + Y_{\mathrm{A3}} \cos3\lpare{{2\pi - \gammaA}} \\&+ Y_{\mathrm{A5}} \cos5\lpare{{2\pi - \gammaA}} \\
       C_{N} = &N_{\mathrm{A1}} \cos\lpare{{2\pi - \gammaA}} + N_{\mathrm{A3}} \cos3\lpare{{2\pi - \gammaA}} \\&+ N_{\mathrm{A5}} \cos5\lpare{{2\pi - \gammaA}} \\
    \end{split}  
\end{equation}
where $X_{\mathrm{A\dots}}, Y_{\mathrm{A\dots}}, N_{\mathrm{A\dots}}$ are empirical coefficients determined from Fujiwara's regression formulae. 

Now, by considering the ship as a rigid body, the rigid‑body kinetics can be summarized in a vector‑matrix formulation \cite{handbook_fossen_2011} but neglecting the nonlinear damping, and Coriolis and centripetal forces as shown in \cref{eqn: kinetics}, then solved:
\begin{equation}\label{eqn: kinetics}
         M\boldsymbol{\dot{x}} = \tau 
\end{equation}
where $M$ is the mass and inertia matrix, $\dot{\boldsymbol{x}}$ is the time derivative of the velocity vector, $\boldsymbol{x} \equiv [\uvelo, \vm, r]^{\intercal} \in \mathbb{R}^3$ and $\tau$ represents the total external forces and moments (such as hydrodynamic, motion-induced, and wind loads) vector acting on the ship and can be decomposed as detailed below \cref{eqn: tau}:

\begin{equation}\label{eqn: tau}
    \tau = F(\boldsymbol{x}) + G(\boldsymbol{u}) + F_{\mathrm{w}}(\uvelo,\vm, \psi, \UT, \gammaT)
\end{equation}
where $F(\boldsymbol{x})$ denotes the motion‑induced hydrodynamic forces and moment, $G(\boldsymbol{u})$ denotes the control forces and moment generated by the rudders, propeller, and bow thruster, with the control input vector defined as, $\boldsymbol{u} \equiv [\deltaptilde, \deltastilde, \ntilde, \iBTtilde]^{\intercal} \in \mathbb{R}^4$, and $F_{\mathrm{w}}$ represents the wind‑induced forces and moments, which is a function of $\uvelo$, $\vm$, $\psi$, and true wind speed ($\UT$) and direction ($\gammaT$).

Terms in \cref{eqn: kinetics} and \cref{eqn: tau} can be obtained from \cref{eq: linear equations of motion}, \cref{eqn: final taylor expansion} and \cref{eqn: air force and moments} as follows:

\begin{equation} \label{eq: final matrix and equation}
    \begin{split}
     M\dot{\boldsymbol{x}} &= \begin{pmatrix} m+m_x & 0 &  0 \\ 0 & m+m_y & 0 \\  0 & 0 & I_{zz}+J_{zz} \end{pmatrix} \begin{pmatrix} \uvelodot \\ \vmdot \\ \rdot \end{pmatrix}\\\\
     F(\boldsymbol{x}) &= \begin{pmatrix} X_{\uvelo} & 0 & 0\\ 0 & Y_{\vm} & Y_{r}\\ 0 & N_{\vm} & N_{r}\end{pmatrix} \begin{pmatrix} \uvelo \\ \vm \\ r \end{pmatrix}\\\\
     G(\boldsymbol{u}) &= \begin{pmatrix} X_{\deltaptilde} & X_{\deltastilde} & X_{\ntilde}& 0 \\ Y_{\deltaptilde} & Y_{\deltastilde} & Y_{\ntilde}& Y_{\iBTtilde}\\ N_{\deltaptilde} & N_{\deltastilde} & N_{\ntilde}& N_{\iBTtilde}\end{pmatrix} \begin{pmatrix} \deltaptilde \\ \deltastilde \\ \ntilde \\ \iBTtilde \end{pmatrix} \\\\
     F_{\mathrm{w}} &= (X_{\mathrm{air}}, Y_{\mathrm{air}}, N_{\mathrm{air}})^\intercal
    \end{split}
\end{equation}

\subsection{Data Curation} \label{sec: data curation}
The details of original operational data are detailed in a previous study by the authors \cite{Mwange2025}. As mentioned in that study, data from two ports was known to involve the usage of tugboats or anchors during the last stage of berthing. The time when the anchor or tugboats were engaged was not recorded in real-time, and therefore, for accurate representation of actual ship dynamics, this data was excluded from the data used in this study, leaving a total of 94 files out of the original 153 files, and the distribution across ports is shown in \cref{tab: ports data information}. 
\begin{table}[H]
        \captionsetup{skip=0pt,singlelinecheck=off, justification=raggedright}
        \caption{Breakdown of log data for each port in the \\ data set. }
        \begin{tabular}{p{0.72cm}|p{0.70cm}p{0.72cm}p{1.05cm}|p{0.70cm}p{0.72cm}p{1.0cm}}
            \toprule
            \multicolumn{1}{c}{Port}   & \multicolumn{3}{c}{Berthing} & \multicolumn{3}{c}{Unberthing} \\\\
            & No. of log data & Original Time (s) & Truncated Time (s) & No. of log data & Original Time (s) & Truncated Time (s) \\
            \midrule
            Port 1 & 19 & 30970 & 17688     &     17 & 20668 & 11589 \\
            Port 2 & 9 &  16065 & 10309    &      8 & 7663 & 4041 \\
            Port 3 & 15 &  23654 & 13201   & 15 &    17422 & 9488 \\
            Port 4 & 4 &  7802  & 635 & 4     &  6148 & 2804\\
            Port 5  & 2 &  3168 & 2087        & 1 &  923 & 469\\
            \bottomrule 
            Total        & 49 & 81659 & 43920 & 45 & 52824 & 28391 \\
        \end{tabular}
    \label{tab: ports data information}
\end{table}
\vspace{1em}

The berthing patterns of Ports 1 - 5 are detailed in \cref{tab: berthing pattern for each port}. 
\begin{table}[H]
    \captionsetup{skip=0pt,singlelinecheck=off, justification=raggedright}
    \caption{Berthing patterns at each port.}
        \begin{tabular}{p{2.5cm} 
        p{5.0cm}}
            \toprule
            Port   & Berthing style\\
            \midrule
            Port 1 & Head-in, STBD moored\\
            Port 2 & $180\degree$ turn to STBD, PORT moored \\
            Port 3 & $90\degree$ turn to STBD, PORT moored \\
            Port 4 & $180\degree$ turn to PORT, STBD moored\\
            Port 5 & $180\degree$ turn to PORT, STBD moored\\
            \bottomrule 
        \end{tabular}
    \label{tab: berthing pattern for each port}
\end{table}
\vspace{1em}

Additionally, the study \cite{Mwange2025} highlighted that there is a high correlation between state variables such as $r$ and $v_{\mathrm{m}}$, and $r$ and drift angle ($\beta$), especially in high-speed regions. The study \cite{Mwange2025} also concluded that the low-speed maneuvering region can be defined as $U \leq 2 \; \mathrm{knots}$. Consequently, the data used in this study was truncated at $U \leq 5 \; \mathrm{knots}$. This truncation effectively excluded a significant portion of data characterized by high speeds and predominantly straight-line motion, thereby establishing a more balanced representation of straight-line and turning maneuvers in the data. 
Moreover, it was noted that there exist differences in maneuvering dynamics during berthing and unberthing \cite{Mwange2025}; consequently, it is recommended to perform identification of model parameters for berthing and unberthing separately. This study presents model parameters for berthing motion only. 

\begin{figure}[htbp]
    \begin{minipage}[htbp]{0.52\columnwidth}
        \centering
        \includegraphics[keepaspectratio, width = 1.0\hsize]{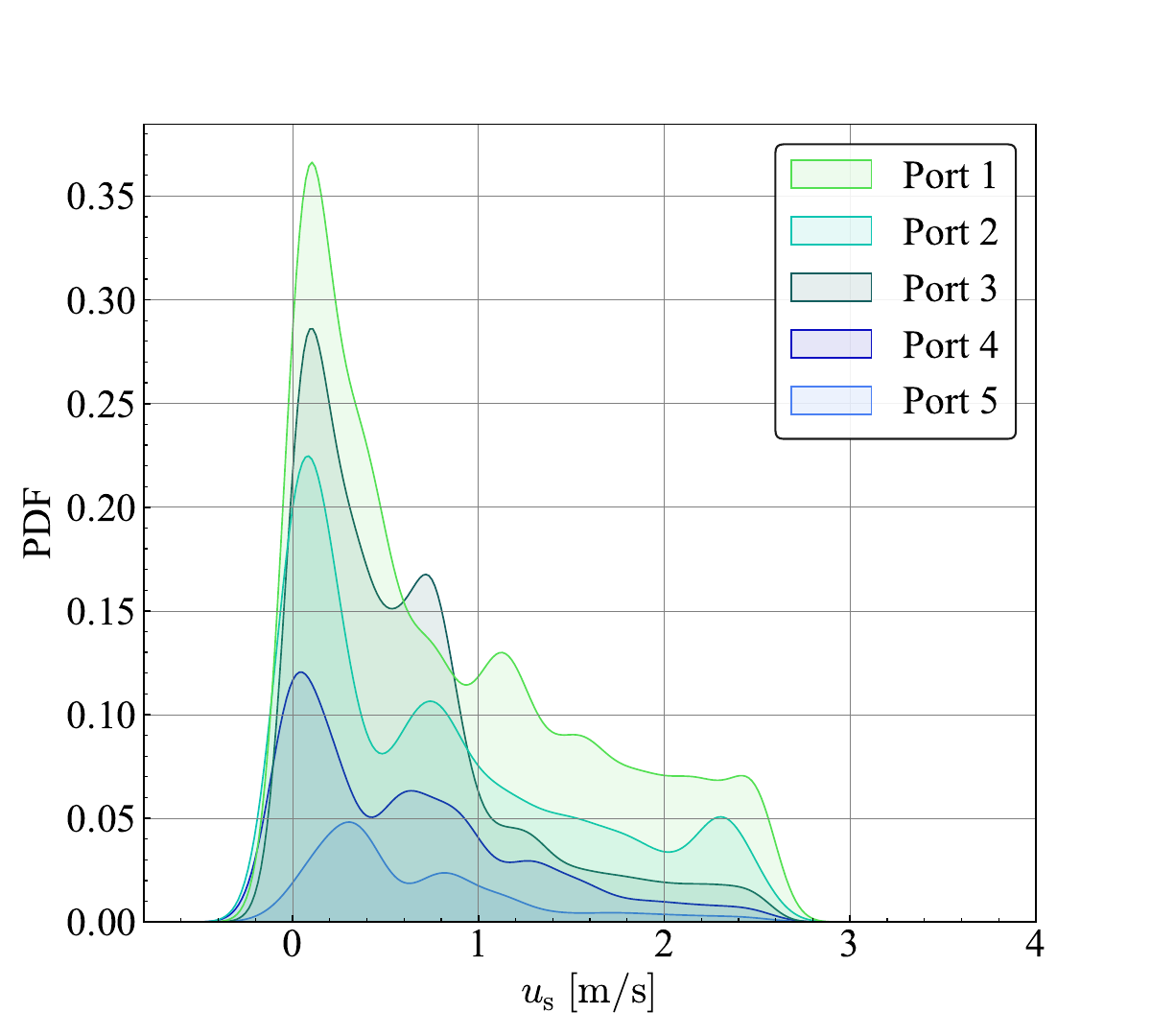}
        \subcaption{Distribution of $u \; [\mathrm{m/s}]$.}
        \label{fig: distribution of u [m/s]}
    \end{minipage}%
    \begin{minipage}[htbp]{0.52\columnwidth}
        \centering
        \includegraphics[keepaspectratio, width = 1.0\hsize]{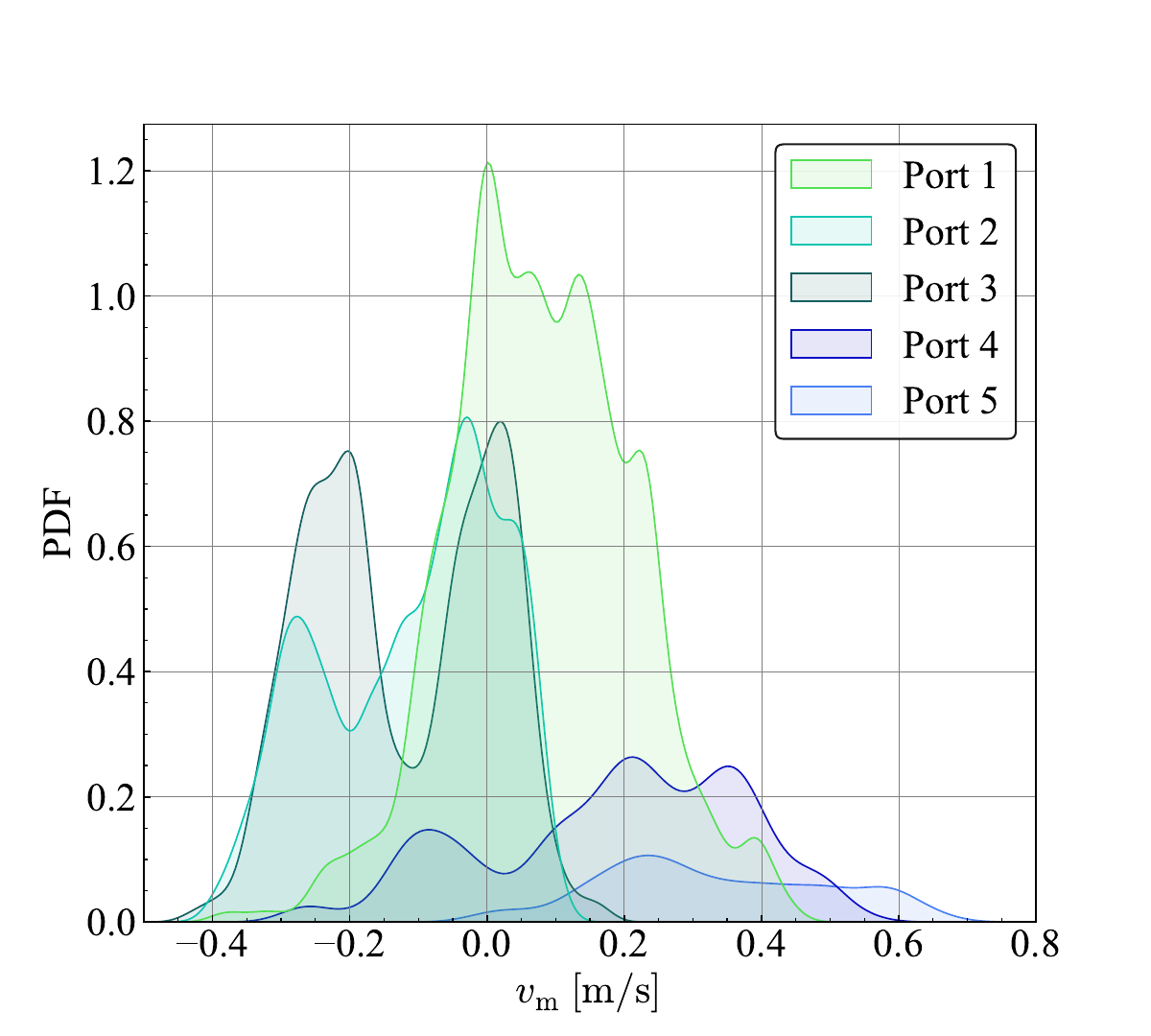}
        \subcaption{Distribution of $v_{\mathrm{m}} \; [\mathrm{m/s}]$.}
        \label{fig: distribution of v_m [m/s]}
    \end{minipage}
    \begin{minipage}[htbp]{0.52\columnwidth}
        \centering
        \includegraphics[keepaspectratio, width = 1.0\hsize]{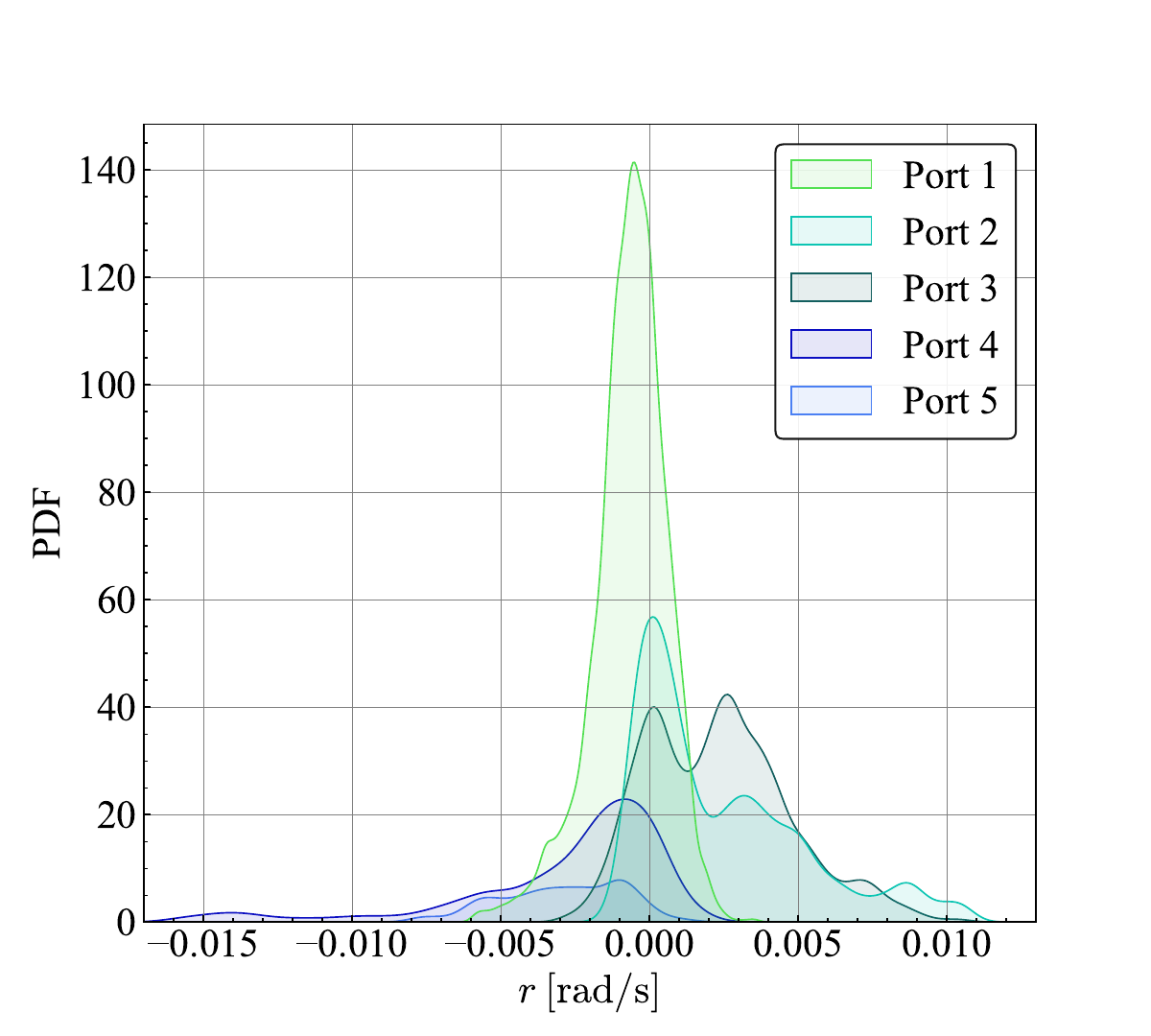}
        \subcaption{Distribution of $r \; [\mathrm{rad/s}]$.}
        \label{fig: distribution of r [rad/s]}
    \end{minipage}%
    \begin{minipage}[htbp]{0.52\columnwidth}
        \centering
        \includegraphics[keepaspectratio, width = 1.0\hsize]{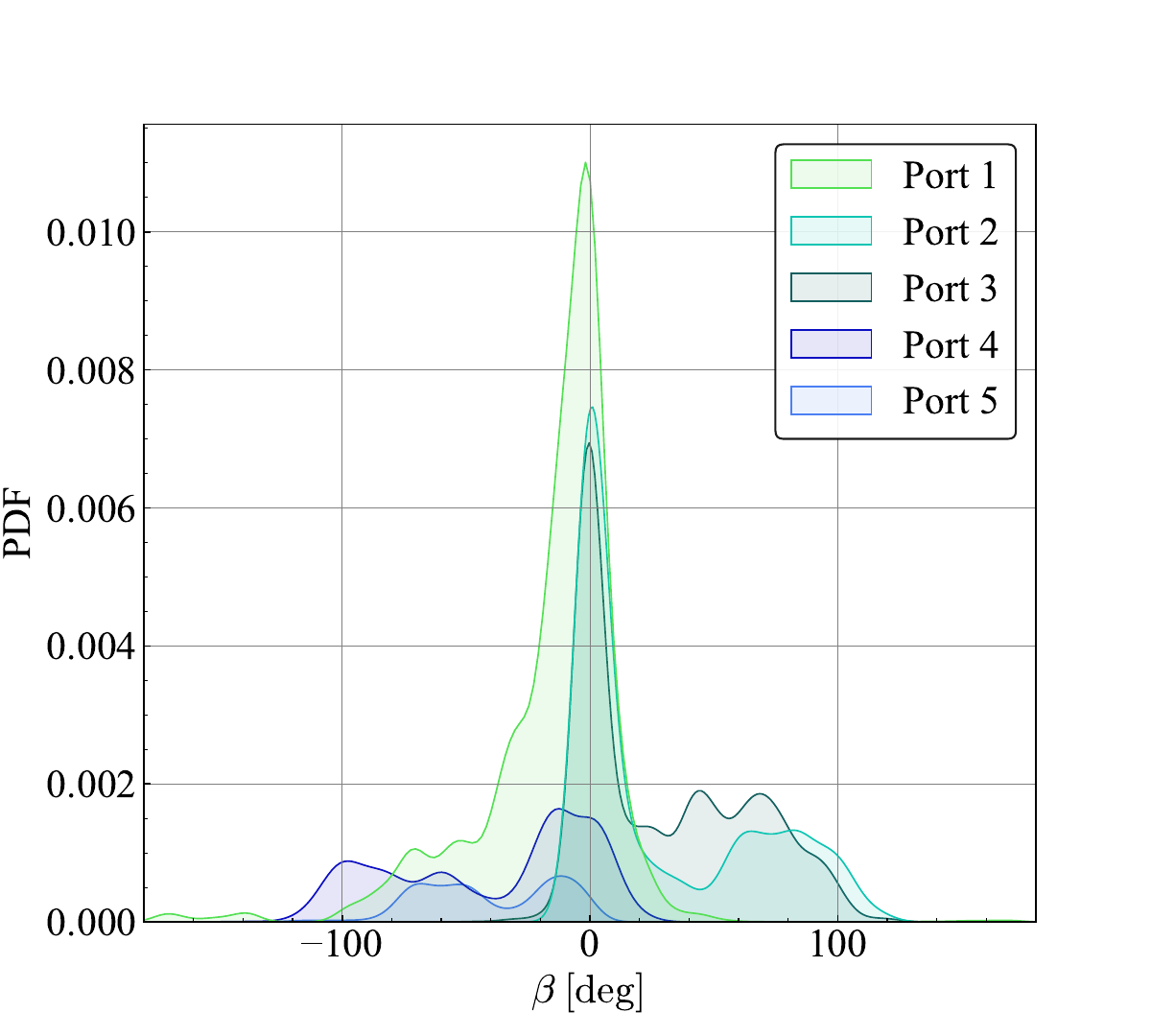}
        \subcaption{Distribution of $\beta \; [\mathrm{deg}]$.}
        \label{fig: distribution of beta [deg]}
    \end{minipage}
    \begin{minipage}[htbp]{0.52\columnwidth}
        \centering
        \includegraphics[keepaspectratio, width = 1.0\hsize]{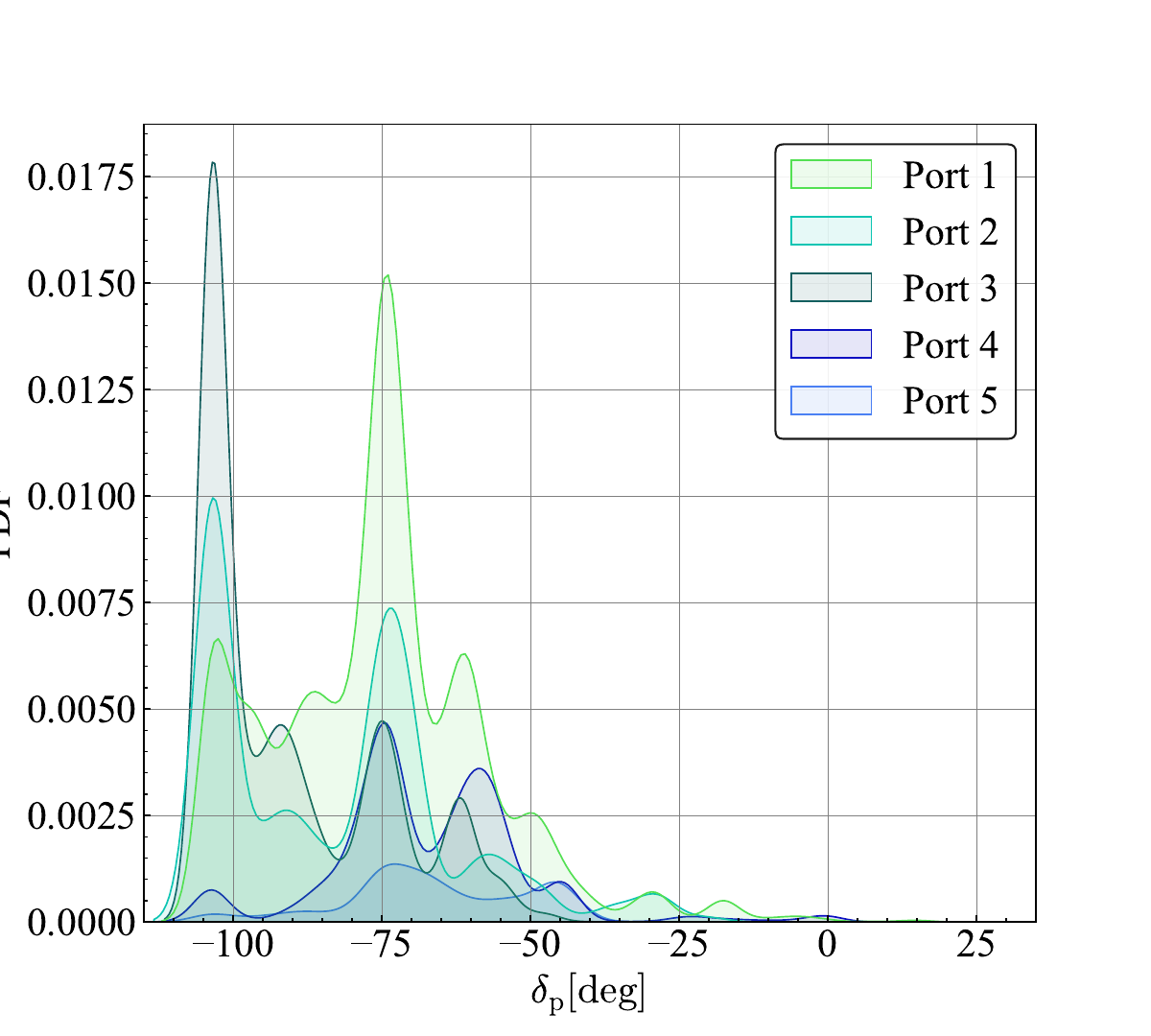}
        \subcaption{Distribution of $\delta_{\mathrm
        p} \; [\mathrm{deg}]$.}
        \label{fig: distribution of delta_p [deg}
    \end{minipage}%
    \begin{minipage}[htbp]{0.52\columnwidth}
        \centering
        \includegraphics[keepaspectratio, width = 1.0\hsize]{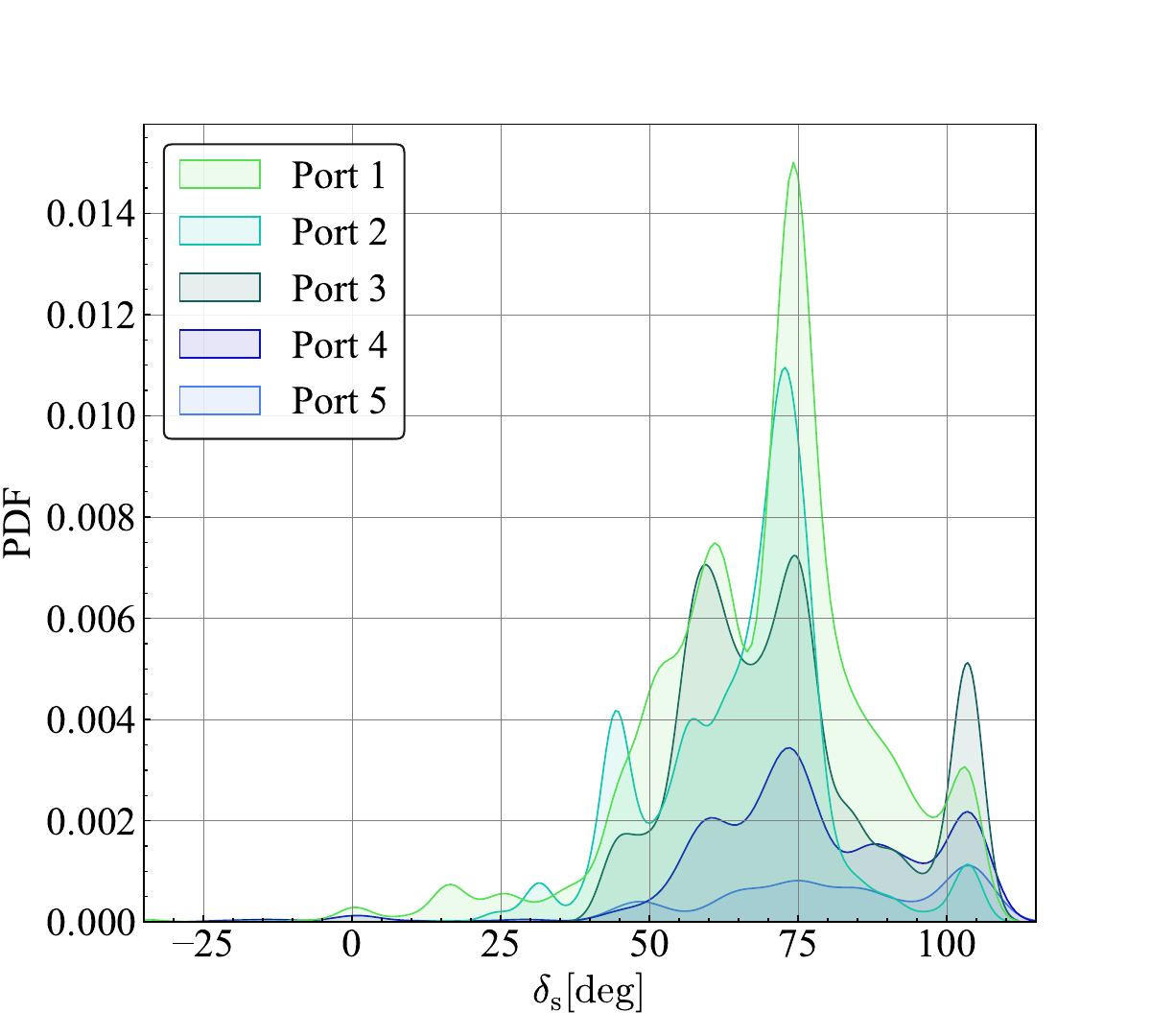}
        \subcaption{Distribution of $\delta_{\mathrm
        s} \; [\mathrm{deg}]$.}
        \label{fig: distribution of delta_s [deg}
    \end{minipage}
    \begin{minipage}[htbp]{0.52\columnwidth}
        \centering
        \includegraphics[keepaspectratio, width = 1.0\hsize]{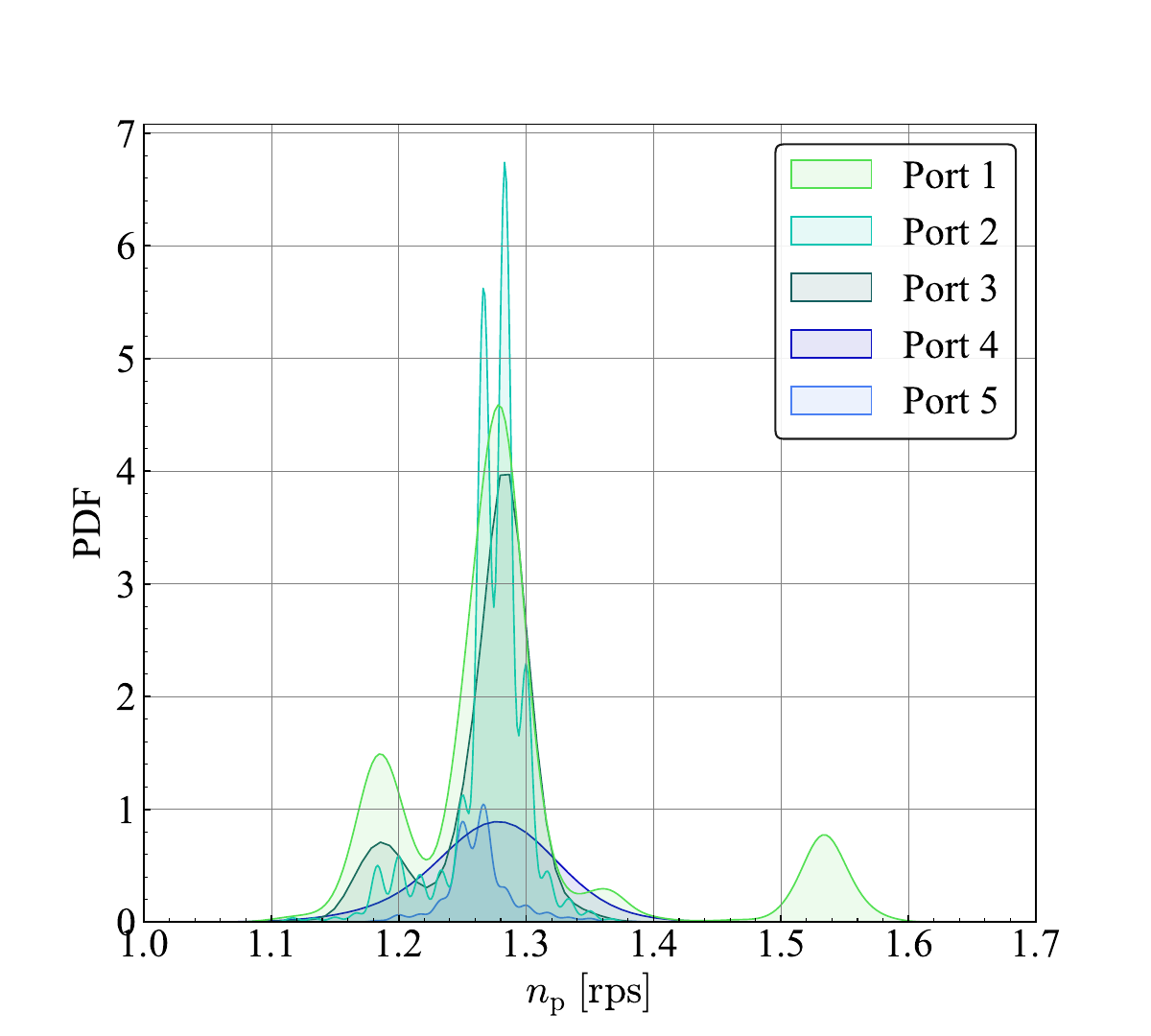}
        \subcaption{Distribution of $n_{\mathrm{p}} \; [\mathrm{rps}]$.}
        \label{fig: distribution of n_p [rps]}
    \end{minipage}%
    \begin{minipage}[htbp]{0.52\columnwidth}
        \centering
        \includegraphics[keepaspectratio, width = 1.0\hsize]{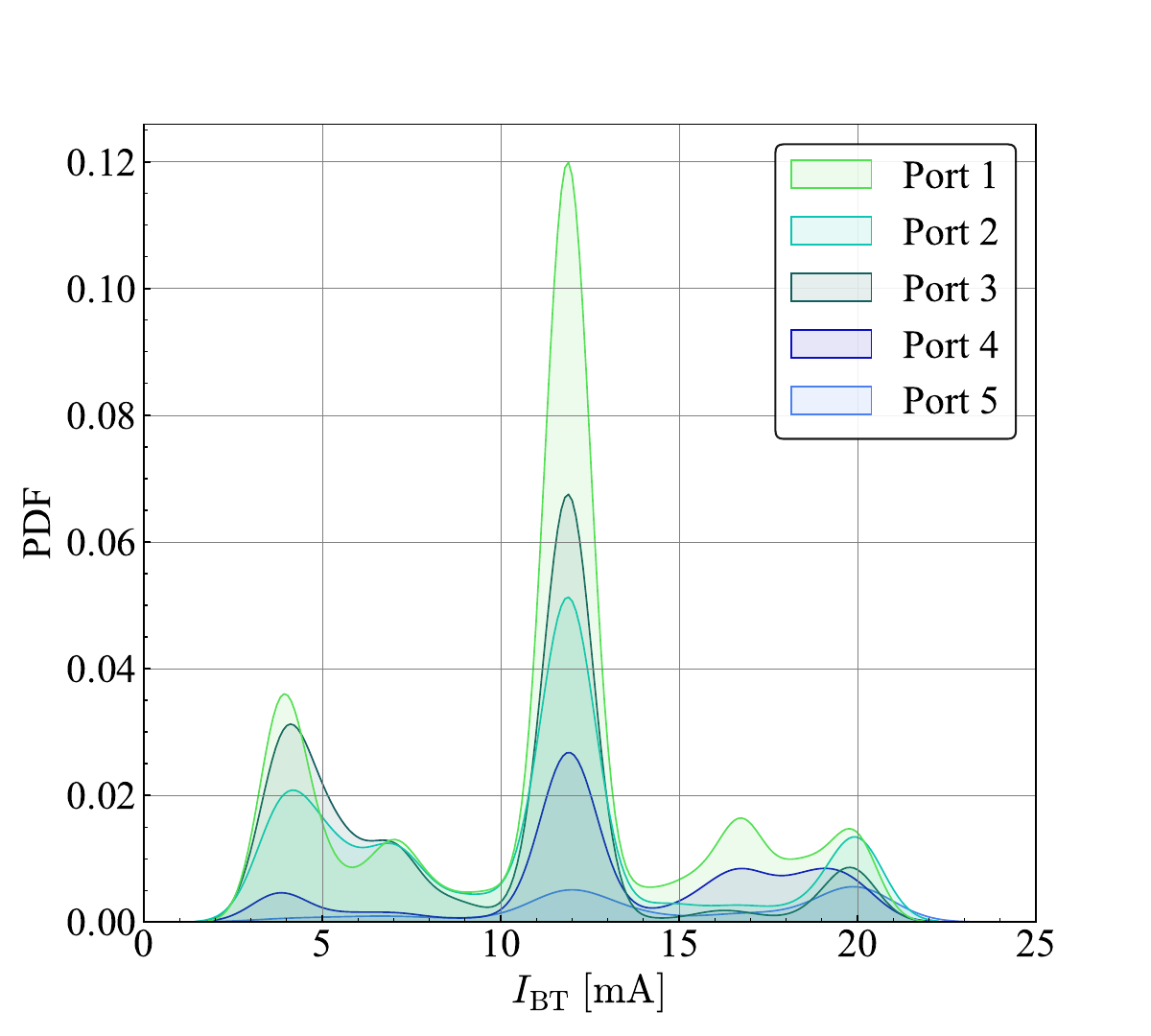}
        \subcaption{Distribution of $I_{\mathrm{BT}} \; [\mathrm{mA}]$.}
        \label{fig: distribution of I_BT [mA]}
    \end{minipage}
\caption{Distribution of state variables and control inputs across Ports 1 - 5.}
\label{fig: kde_gen_distribution}
\end{figure}

\begin{figure}[htbp]
    \begin{minipage}[htbp]{0.52\columnwidth}
        \centering
        \includegraphics[keepaspectratio, width = 1.0\hsize]{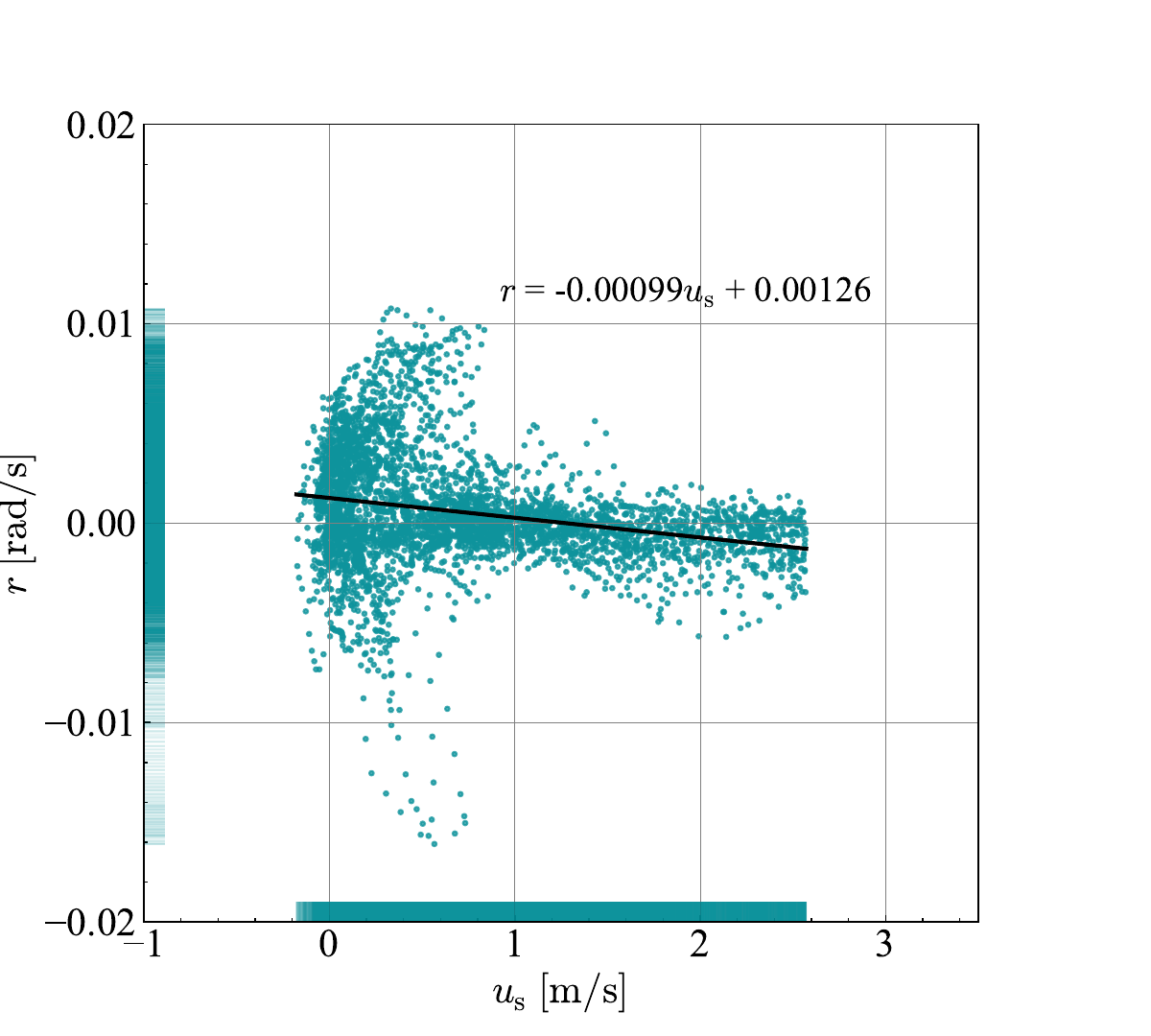}
        \subcaption{Correlation between $u$ and $r$. }
        \label{fig: correlation between u and r }
    \end{minipage}%
    \begin{minipage}[ht]{0.52\columnwidth}
        \centering
        \includegraphics[keepaspectratio, width = 1.0\hsize]{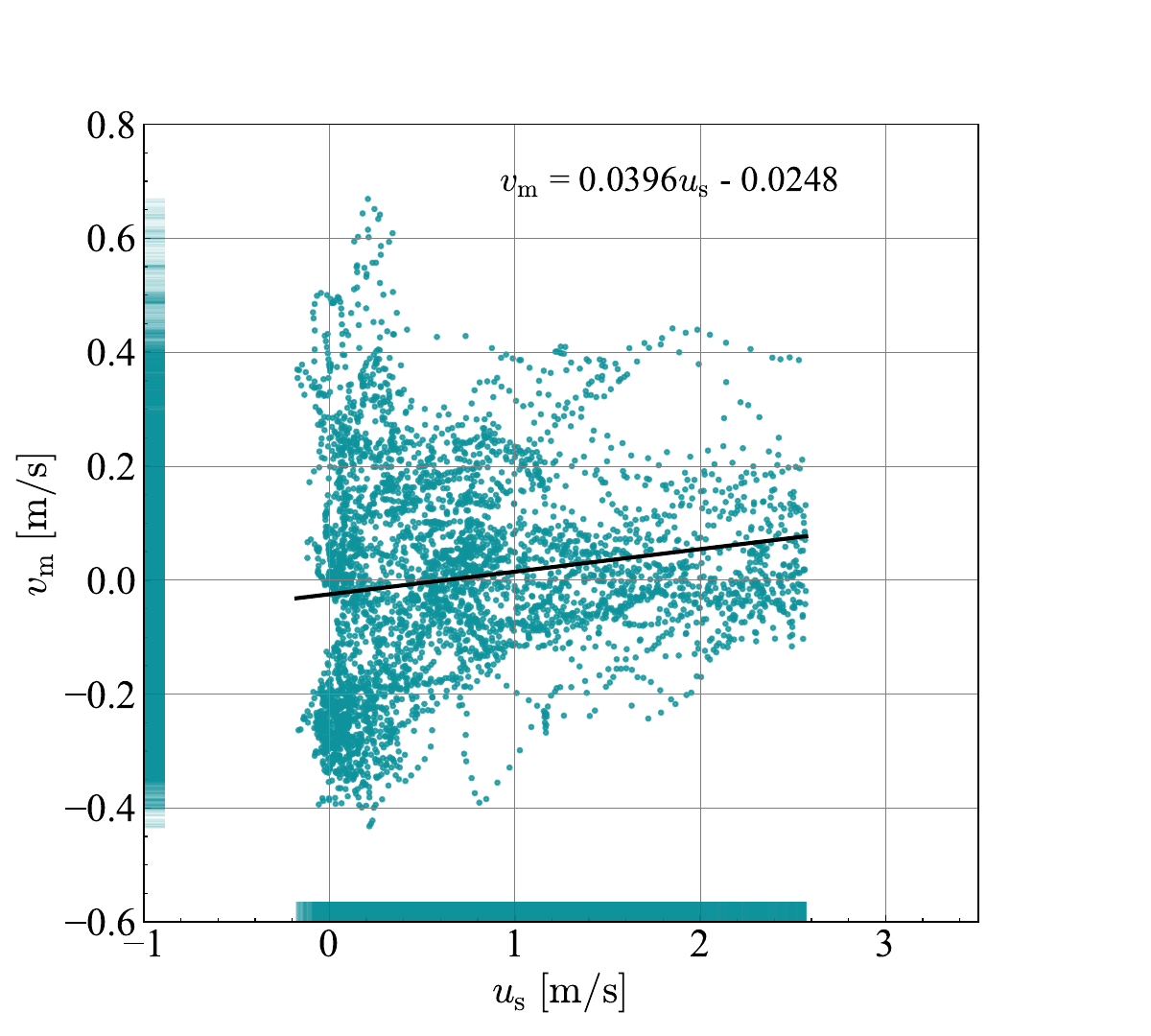}
        \subcaption{Correlation between $u$ and $\vm$. }
        \label{fig: correlation between u and vm }
    \end{minipage}
    \begin{minipage}[ht]{0.52\columnwidth}
        \centering
        \includegraphics[keepaspectratio, width = 1.0\hsize]{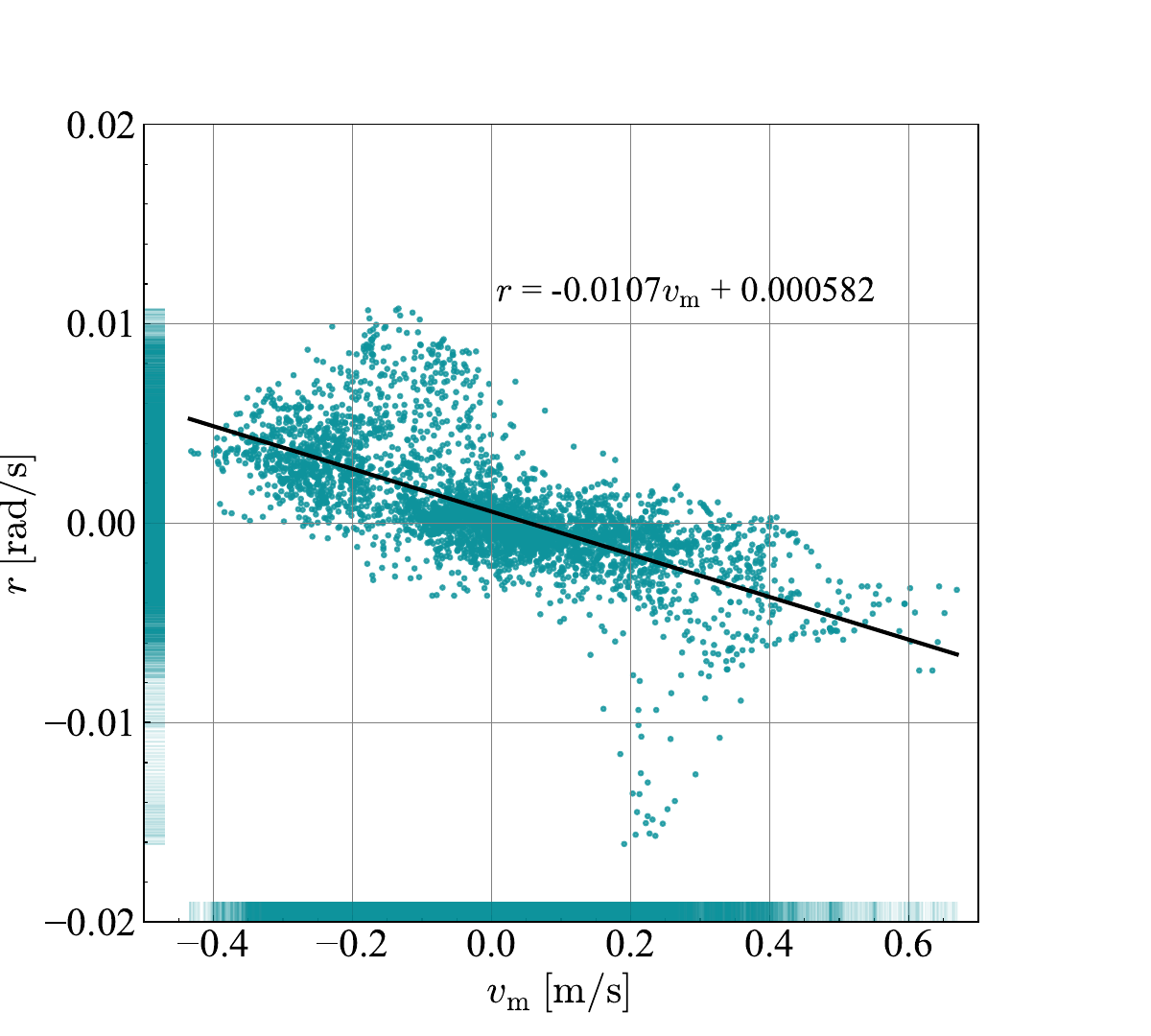}
        \subcaption{Correlation between $\vm$ and $r$. }
        \label{fig: correlation between vm and r }
    \end{minipage}%
    \begin{minipage}[ht]{0.52\columnwidth}
        \centering
        \includegraphics[keepaspectratio, width = 1.0\hsize]{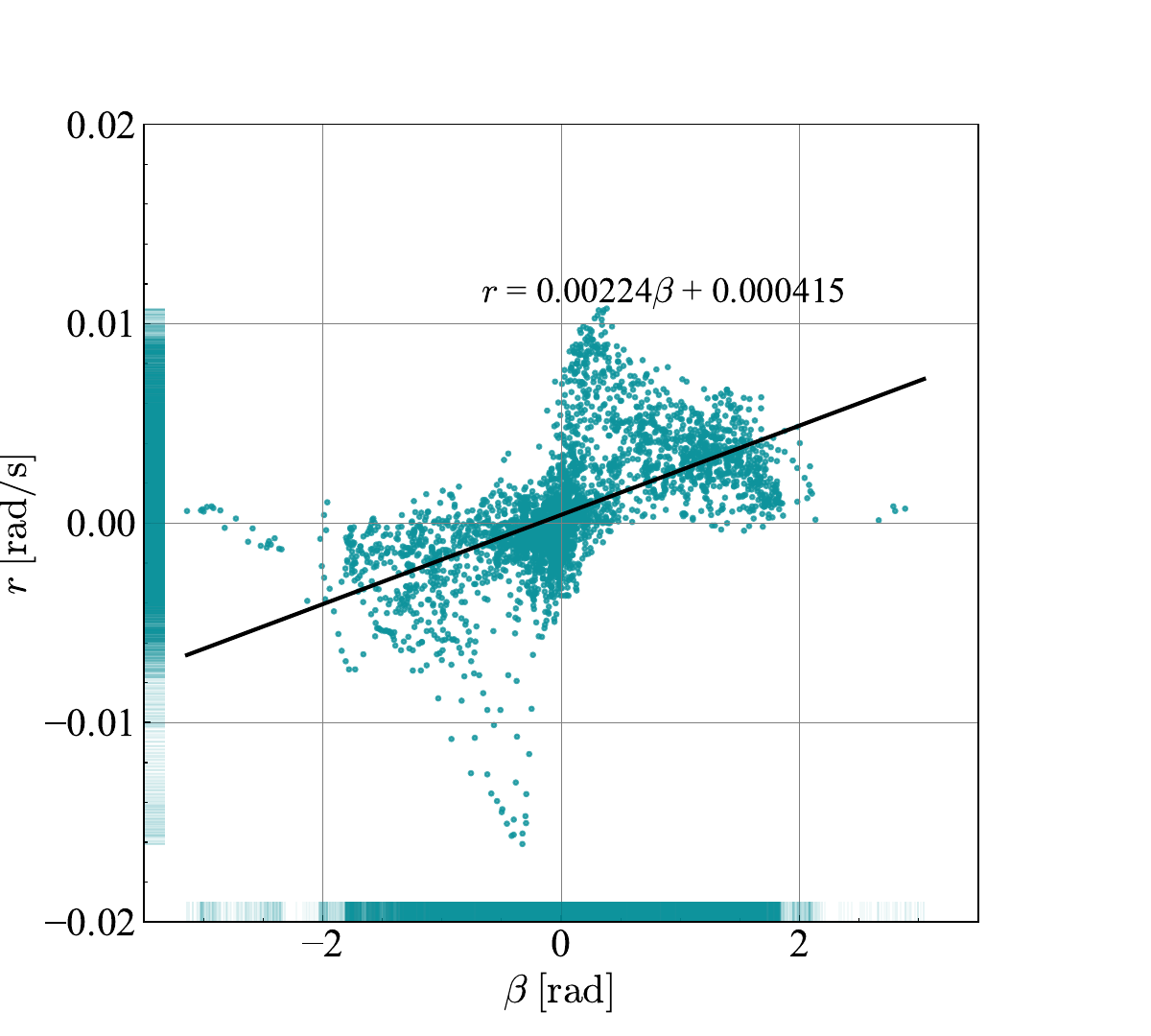}
        \subcaption{Correlation between $\beta$ and $r$. }
        \label{fig: correlation between beta and r }
    \end{minipage}
    \begin{minipage}[htbp]{0.52\columnwidth}
        \centering
        \includegraphics[keepaspectratio, width = 1.0\hsize]{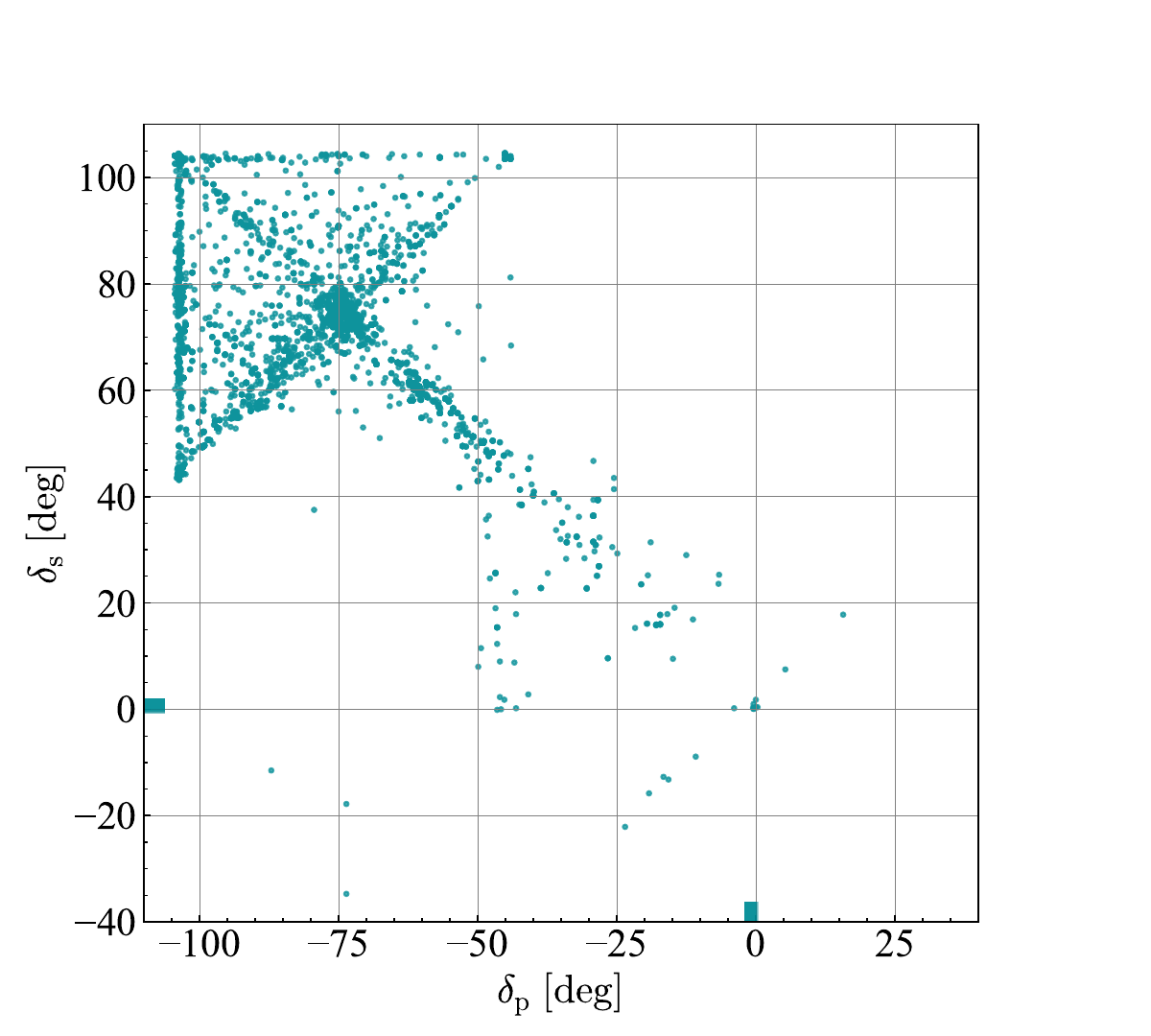}
        \subcaption{Correlation between $\deltap$ and $\deltas$. }
        \label{fig: correlation between delta_p and delta_s }
    \end{minipage}%
    \begin{minipage}[ht]{0.52\columnwidth}
        \centering
        \includegraphics[keepaspectratio, width = 1.0\hsize]{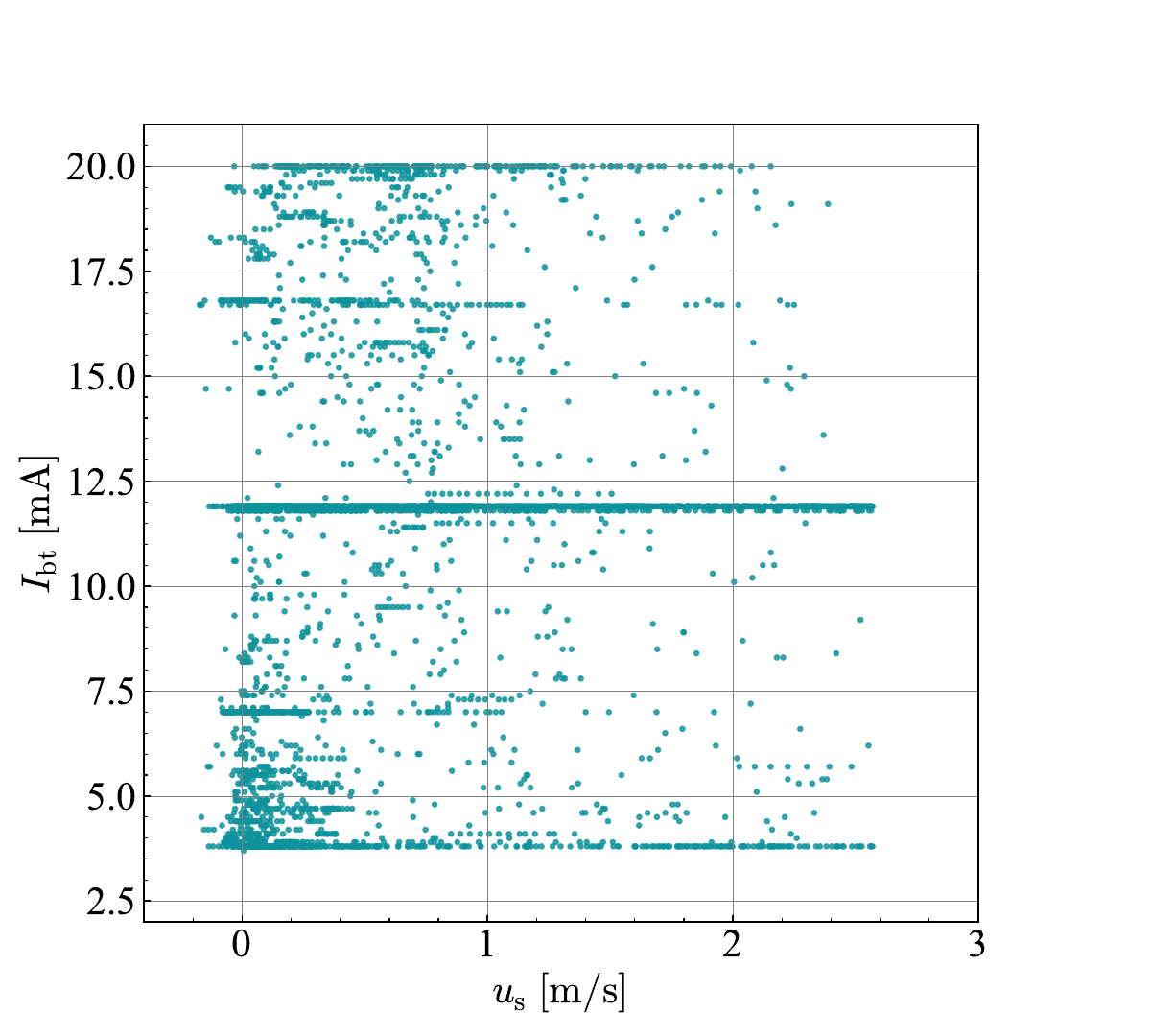}
        \subcaption{Correlation between $u$ and $I_{\mathrm{BT}}$. }
        \label{fig: correlation between u and I_BT }
    \end{minipage}

\caption{ Correlation between state variables and control inputs in the whole data set.}
\label{fig: correlation plots}
\end{figure}

The overall distribution of the data used in this study ($\mathcal{D}$) across each port is shown in \cref{fig: kde_gen_distribution}. Further, \cref{fig: correlation plots} displays correlations between essential state and control variables. The correlation plots were prepared with data resampled at 0.1Hz, whereas the numerical correlation values were derived from the complete dataset without resampling.

\subsection{Optimization Problem} \label{sec: optimization problem}
\begin{figure}[htbp]
    \begin{minipage}[htbp]{0.52\columnwidth}
        \centering
        \includegraphics[keepaspectratio, width = 1.0\hsize]{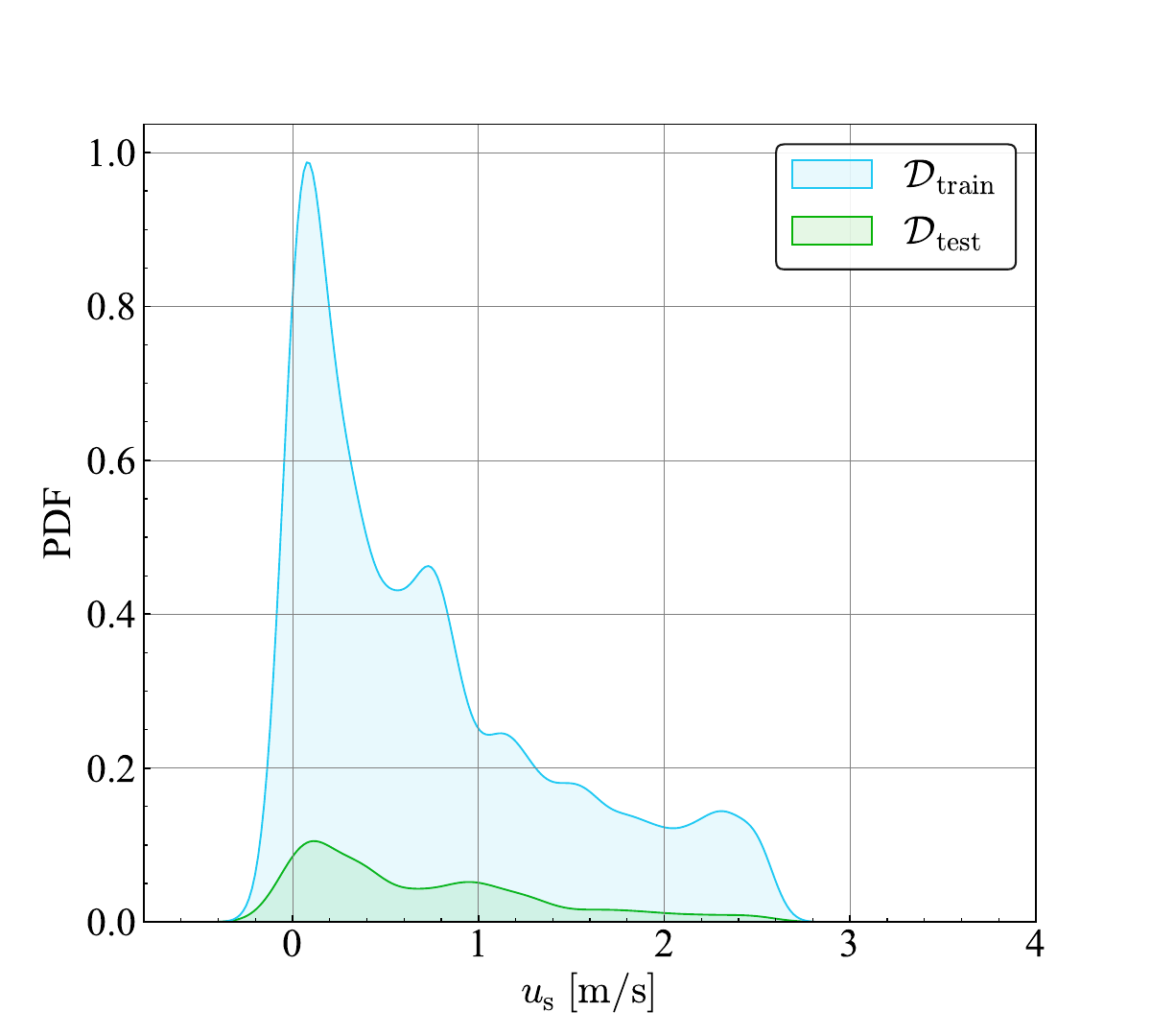}
        \subcaption{Distribution of $u \; [\mathrm{m/s}]$.}
        \label{fig: distribution of u [m/s] in the training and testing datasets.}
    \end{minipage}%
    \begin{minipage}[ht]{0.52\columnwidth}
        \centering
        \includegraphics[keepaspectratio, width = 1.0\hsize]{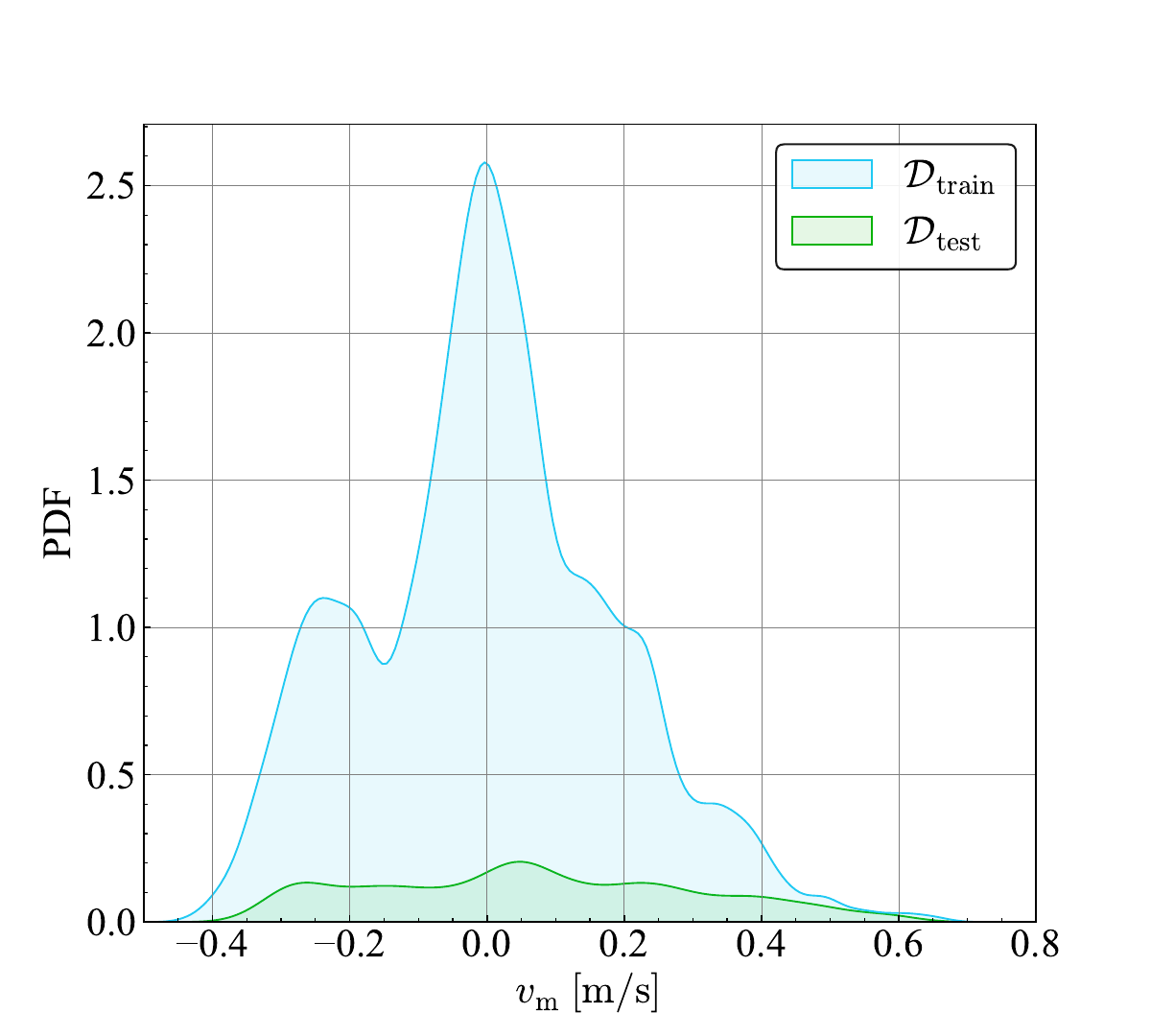}
        \subcaption{Distribution of $\vm \; [\mathrm{m/s}]$.}
        \label{fig: distribution of v_m [m/s] in the training and testing datasets.}
    \end{minipage}
    \begin{minipage}[ht]{0.52\columnwidth}
        \centering
        \includegraphics[keepaspectratio, width = 1.0\hsize]{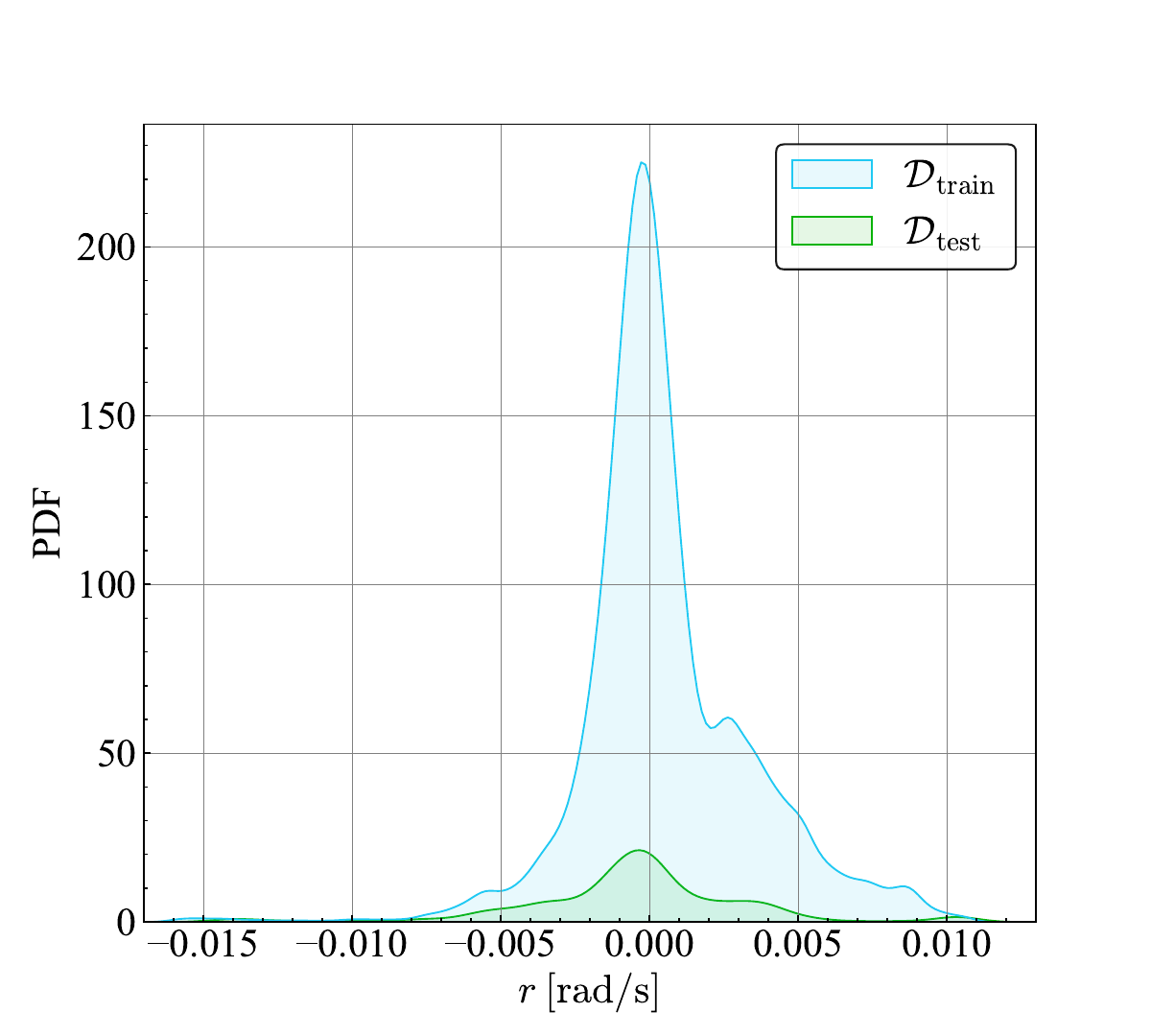}
        \subcaption{Distribution of $r\; [\mathrm{rad/s}]$.}
        \label{fig: distribution of r rad/s] in the training and testing datasets.}
    \end{minipage}%
    \begin{minipage}[ht]{0.52\columnwidth}
        \centering
        \includegraphics[keepaspectratio, width = 1.0\hsize]{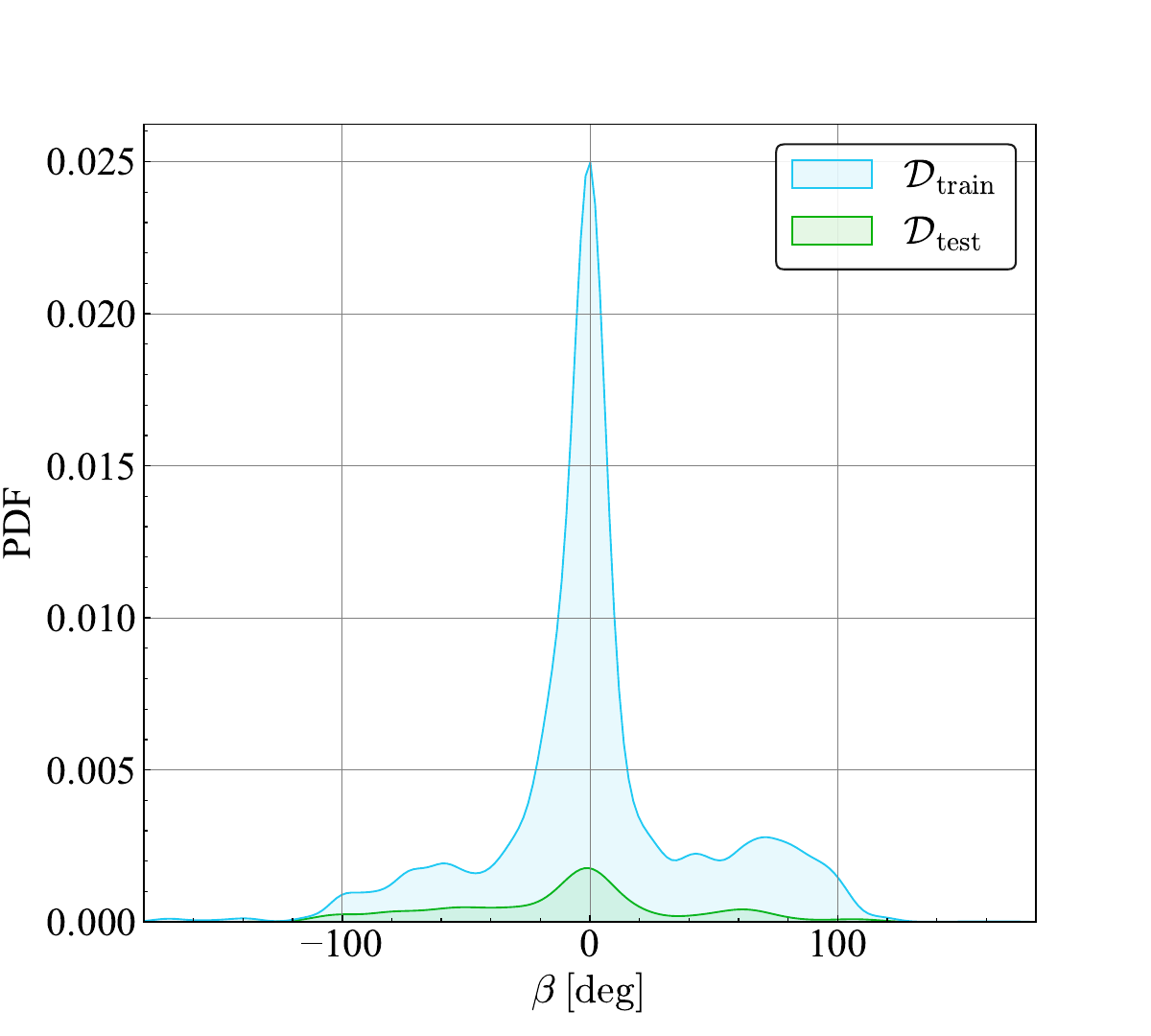}
        \subcaption{Distribution of $\beta\; [\mathrm{deg}]$.}
        \label{fig: distribution of beta [deg in the training and testing datasets.}
    \end{minipage}
    \begin{minipage}[htbp]{0.52\columnwidth}
        \centering
        \includegraphics[keepaspectratio, width = 1.0\hsize]{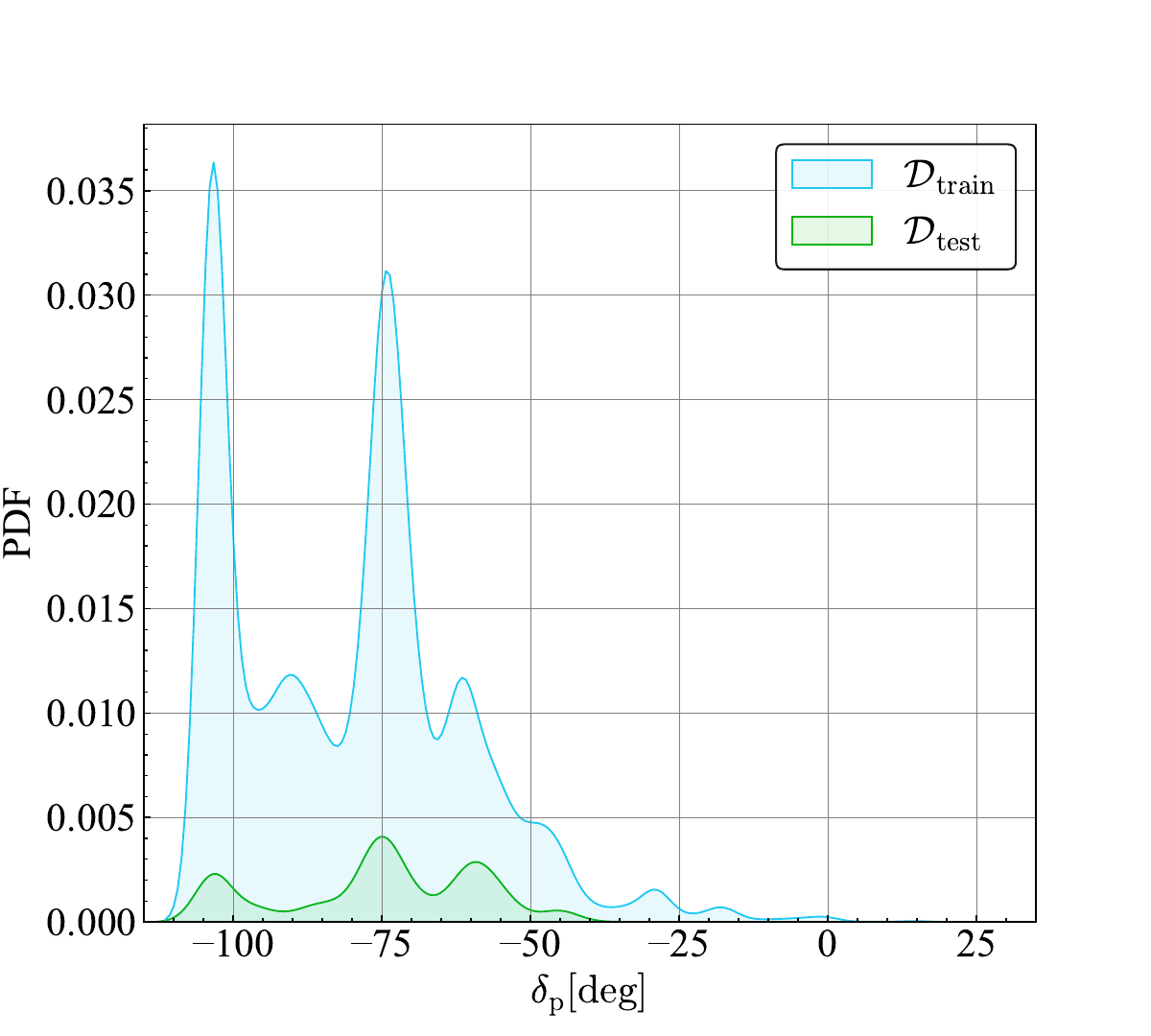}
        \subcaption{Distribution of $\deltap \; [\mathrm{deg}]$.}
        \label{fig: distribution of delta_p [deg] in the training and testing datasets.}
    \end{minipage}%
    \begin{minipage}[ht]{0.52\columnwidth}
        \centering
        \includegraphics[keepaspectratio, width = 1.0\hsize]{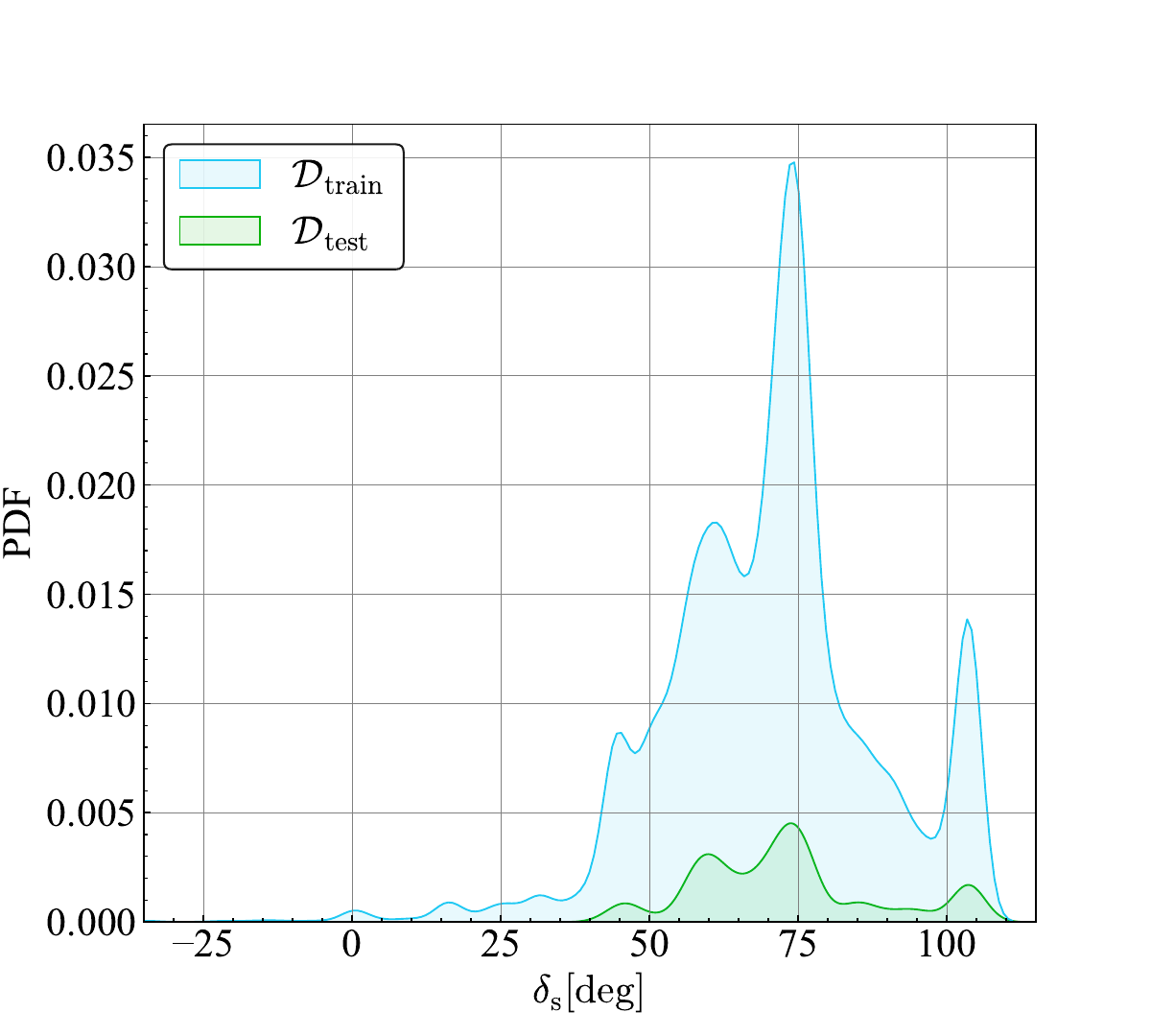}
        \subcaption{Distribution of $\deltas \; [\mathrm{deg}]$.}
        \label{fig: distribution of delta_s [deg] in the training and testing datasets.}
    \end{minipage}
    \begin{minipage}[htbp]{0.52\columnwidth}
        \centering
        \includegraphics[keepaspectratio, width = 1.0\hsize]{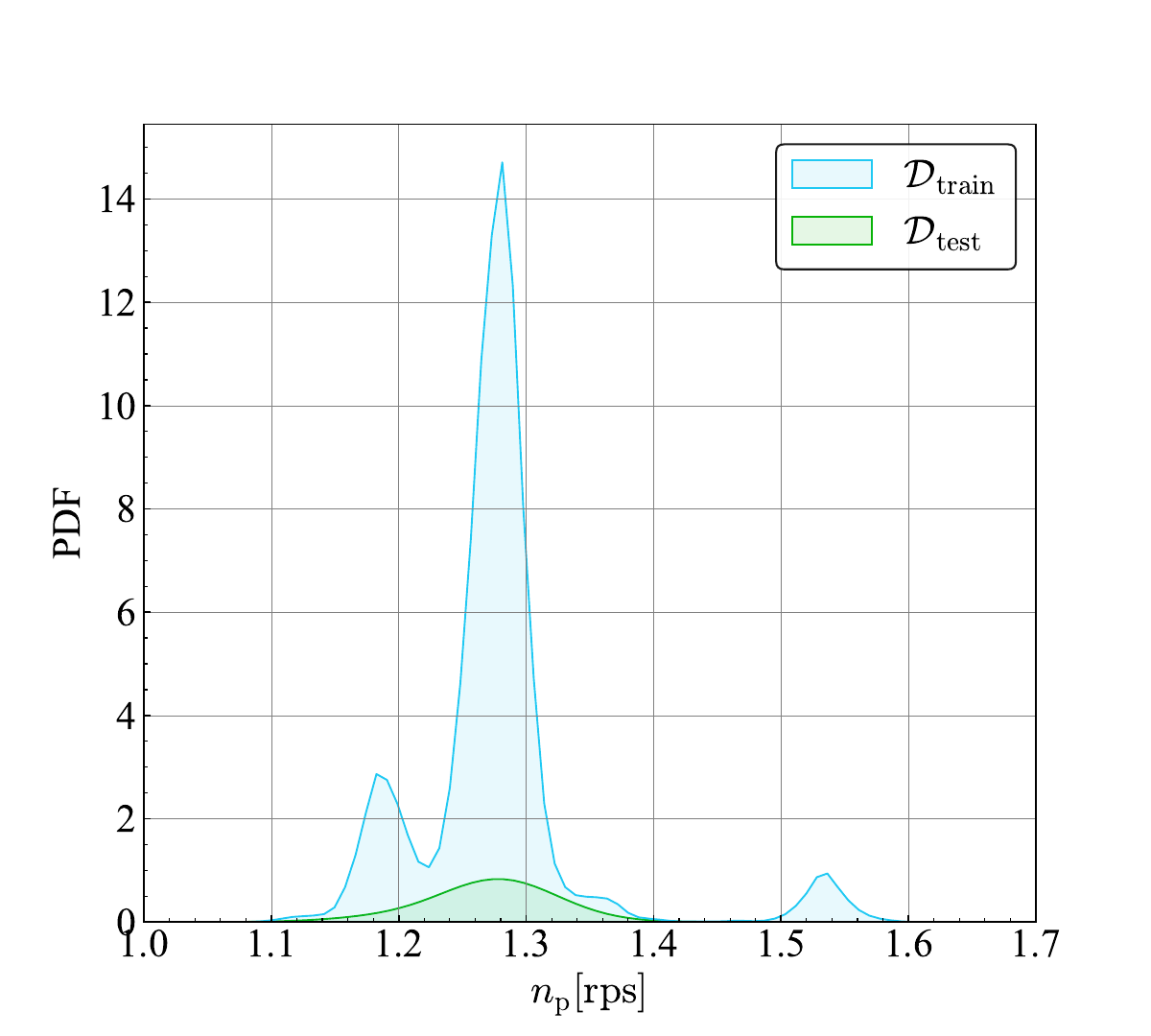}
        \subcaption{Distribution of $n_{\mathrm{p}} \; [\mathrm{rps}]$.}
        \label{fig: distribution of n_p [rps] in the training and testing datasets.}
    \end{minipage}%
    \begin{minipage}[ht]{0.52\columnwidth}
        \centering
        \includegraphics[keepaspectratio, width = 1.0\hsize]{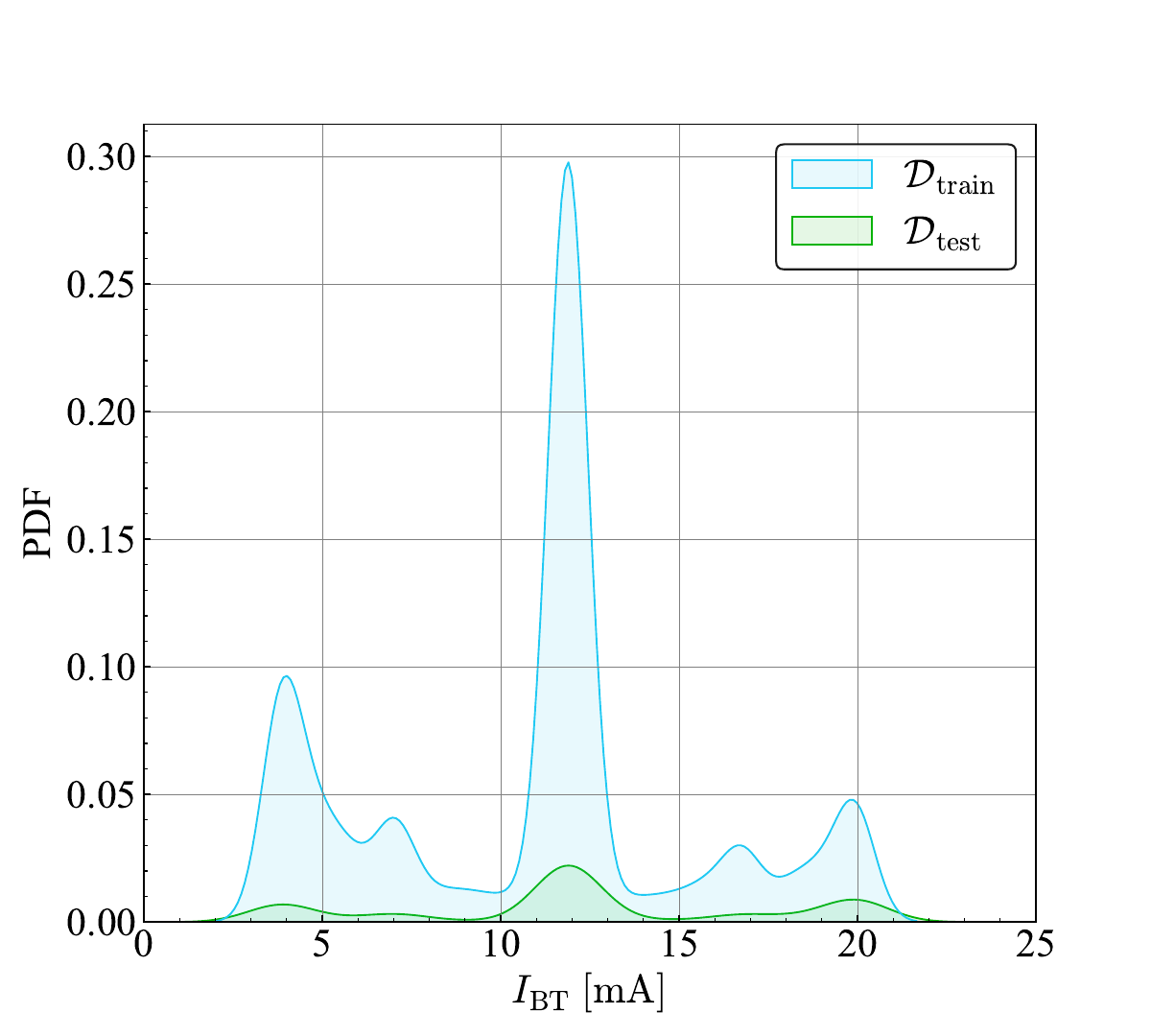}
        \subcaption{Distribution of $I_{\mathrm{BT}} \; [\mathrm{mA}]$.}
        \label{fig: distribution of Ibt in the training and testing datasets.}
    \end{minipage}
\caption{Distribution of state variables and control inputs in the training and testing data sets.}
\label{fig: training and testing data kde}
\end{figure}

As detailed in \cref{sec: data curation}, this study utilizes the curated berthing operations dataset $\mathcal{D}$, consisting of 49 logfiles. For the purpose of system identification and model validation, $\mathcal{D}$ was partitioned into two distinct subsets: a training dataset ($\mathcal{D}_{\mathrm{train}}$) containing 44 logfiles, and a testing dataset ($\mathcal{D}_{\mathrm{test}}$) containing the remaining 5 logfiles (one logfile per port). The relative distributions of state and control variables within both $\mathcal{D}_{\mathrm{train}}$ and $\mathcal{D}_{\mathrm{test}}$ are visualized in \cref{fig: training and testing data kde}. This partition was constructed such that $\mathcal{D}_{\mathrm{test}}$ constitutes a proper subset of $\mathcal{D}_{\mathrm{train}}$, ensuring that the validation data is representative of the states and control variables distribution inherent to the training data.

Let $\Theta$, $\theta$, and $\theta^{*}$ denote the model parameters exploration domain, the model parameters vector and optimal model parameters vector, respectively. The model parameters vector, $\theta$, has 31 parameters and can be summarized as the added mass and inertia in \cref{eq: final matrix and equation}, the hydrodynamic derivatives in \cref{eqn: final taylor expansion}, the rudder hover angles in \cref{eqn: control dif. hover} and the wind coefficients in \cref{eqn: wind coeffs} as shown below: 
\begin{equation}
\begin{split}
   \theta^{*} \in & \;\Theta \\
   \theta =& \; (\; m_x, m_y, I_{zz}+J_{zz}, X_{\uvelo}, Y_{\vm}, Y_r, N_{\vm}, N_r, X_{\deltaptilde}, X_{\deltastilde}, X_{\ntilde},  \\
   &\;Y_{\deltaptilde} ,Y_{\deltastilde}, Y_{\ntilde}, Y_{\iBTtilde}, N_{\deltaptilde} , N_{\deltastilde}, N_{\ntilde},   N_{\iBTtilde}, \deltaphover, \deltashover,\\
   &\; X_{\mathrm{A0}}, X_{\mathrm{A1}}, X_{\mathrm{A3}}, X_{\mathrm{A5}}, Y_{\mathrm{A1}}, Y_{\mathrm{A3}}, Y_{\mathrm{A5}}, N_{\mathrm{A1}}, N_{\mathrm{A3}}, N_{\mathrm{A5}} \; )
\end{split}
\end{equation}
The exploration ranges for the added mass and inertia parameters ($m_x, m_y$ and $I_{zz} +J_{zz}$) and the coefficients of the Fujiwara regression formula in \cref{eqn: wind coeffs} were determined from empirical data. The manufacturer-specified operational range for the hover rudder angles is $|\deltaphover|, |\deltashover| \in [70^\circ, 80^\circ]$. For the remaining hydrodynamic derivative parameters, the initial exploration bounds were established as $-1 \times 10^{6} \leq \theta_{j} \leq 1 \times 10^{6} $ with the exception of $Y_{\vm}$ and $N_{\vm}$ which based on empirical observation, are negative, and therefore their initial bounds were set to $-1 \times 10^{6} \leq \theta_{j} \leq 0$. The search space for the hydrodynamic derivatives was adaptively expanded based on the results of the computational analysis. Finally, to ensure computational stability and equitable parameter influence, the exploration range for all parameters was standardized to a uniform interval of $\Theta = [-1.0, 1.0 ]$

Now, by considering parameter identification as a constrained optimization problem, the optimal parameter vector $\theta^{*}$  is defined as the one that minimizes the error between the model-simulated states ($q_{\mathrm{sim}}$) and the corresponding measured states on the actual ship ($q_{\mathrm{in}}$), such that:
\begin{equation}
    \theta^{*} = \mathrm{argmin}_{\theta \in \Theta} \; J(\theta; \mathcal{D_{\mathrm{train}}})
\end{equation}
where $J$ is the objective function was proposed by Miyauchi et al. \cite{Miyauchi2022SI} in the form of an L2 norm between the corresponding standardized vectors; ($\hat{q}_{\mathrm{sim}}$) and ($\hat{q}_{\mathrm{in}}$) as follows:
\begin{equation}
         J = \sumstate{\int_{0}^{\tfpla} {\norm{ L }}_{2} ^ {2 } \; \dt }  
\end{equation}
where:
\begin{equation}
    \begin{split}
        \norm{L}_{2} =& \sqrt{\lpare{\qhatinI (t) - \qhatsimI (t)}.\lpare{\qhatinI (t) - \qhatsimI (t)}}\\\\
        \qhatinI (t) =& \; \lpare{\qinI (t) - \mu^{i}_{\mathrm{in}}}/\sigma^{i}_{\mathrm{in}}  \\\\
        \qhatsimI (t) =& \; \lpare{\qsimI(t) - \mu^{i}_{\mathrm{sim}}}/\sigma^{i}_{\mathrm{sim}}     
    \end{split}
\end{equation}
The subscript $i$ denotes the $i_{\mathrm{th}}$ contiguous logfile in $\mathcal{D}_{\mathrm{train}}$ and $N$ is the number of logfiles such that $i = 1, \dots,N $. $\tfpla$ is the time duration of each logfile.  $\mu^{i}$ and $\sigma^{i}$ denote the mean and standard deviation of $q^{i} (t)$. The choice of states included in $q(t)$ varies with the complexity of maneuvers considered in the optimization problem. In this study, $q(t) \equiv [\uvelo(t), \vm(t), r(t) ] ^ \intercal \in \mathbb{R}^3$.

The problem was then solved using the CMA-ES optimization scheme \cite{sakamoto2017modified, maki2020application}.

\section{ Results } \label{sec: results}
The optimal parameters identified for the proposed linear low-speed maneuvering model are summarized in \cref{tab: optimal params}.  

\begin{table}[H] 
\captionsetup{skip=0pt,singlelinecheck=off, justification=raggedright}
    \caption{Optimal model parameters.}

        \begin{tabular}{p{1.1cm} |p{1.7cm}||p{1.1cm} |p{1.7cm}}
            \toprule
            Parameter  & Value & Parameter & Value\\
            \midrule
            $m_x$& 0.0005 & $N_{\deltaptilde}$ & -111648.5007 \\
            $m_y$&0.0001 & $N_{\deltastilde}$ & -107497.1403 \\
            $I_{zz}+J_{zz}$& 0.0041 & $N_{\ntilde}$ & -3.0176 \\
            $X_{\uvelo}$& -10549.9383 & $N_{\iBTtilde}$ & 30741.706 \\
            $Y_{\vm}$& -12710.7039 & $\deltaphover$ & -80.00 \\
            $Y_r$& -9997.126 & $\deltashover$ & 76.61 \\
            $N_{\vm}$& -60434.8145 & $X_{\mathrm{A0}}$ & -2.2778\\
            $N_r$& -15217771.41 & $X_{\mathrm{A1}}$ & -3.9166 \\
            $X_{\deltaptilde}$& -17585.3455 & $X_{\mathrm{A3}}$ & -0.8592 \\
            $X_{\deltastilde}$& 12229.4712& $X_{\mathrm{A5}}$ &  1.8442 \\
            $X_{\ntilde}$& 5627.4898 & $Y_{\mathrm{A1}}$ & 0.5029 \\
            $Y_{\deltaptilde}$& 6816.3598 & $Y_{\mathrm{A3}}$ & 0.1898 \\
            $Y_{\deltastilde}$& 5758.5219 & $Y_{\mathrm{A5}}$ & -0.1528 \\
            $Y_{\ntilde}$& 5161.7409 & $N_{\mathrm{A1}}$ & -0.0106 \\
            $Y_{\iBTtilde}$& 1486.2747 & $N_{\mathrm{A3}}$ & 0.0242 \\
            &&$N_{\mathrm{A5}}$& 0.0126 \\
            \bottomrule 
        \end{tabular}
    \label{tab: optimal params}
\end{table}

\begin{figure}[htbp]
    \centering
    \includegraphics[width = 1.0\columnwidth]{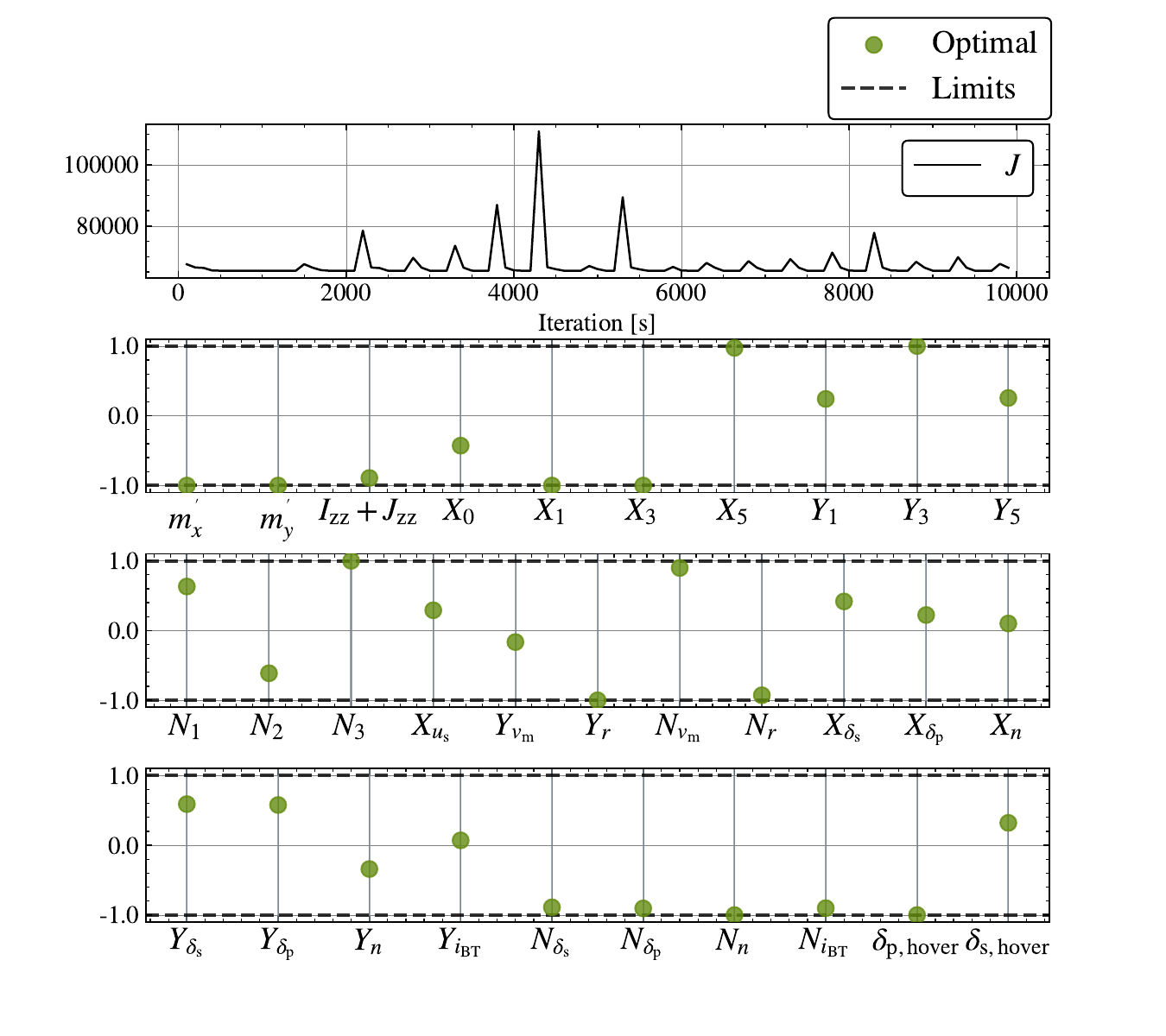}
    \caption{}
    \label{fig: obj func and param limits}
\end{figure}

\cref{fig: obj func and param limits} illustrates the convergence of the objective function during optimization and confirms that the optimal parameter values reside within their prescribed exploration limits. 

\begin{figure}[htbp]
\begin{minipage}[h!]
    {\linewidth}
        \centering
    \includegraphics[keepaspectratio, width = 0.95\hsize]{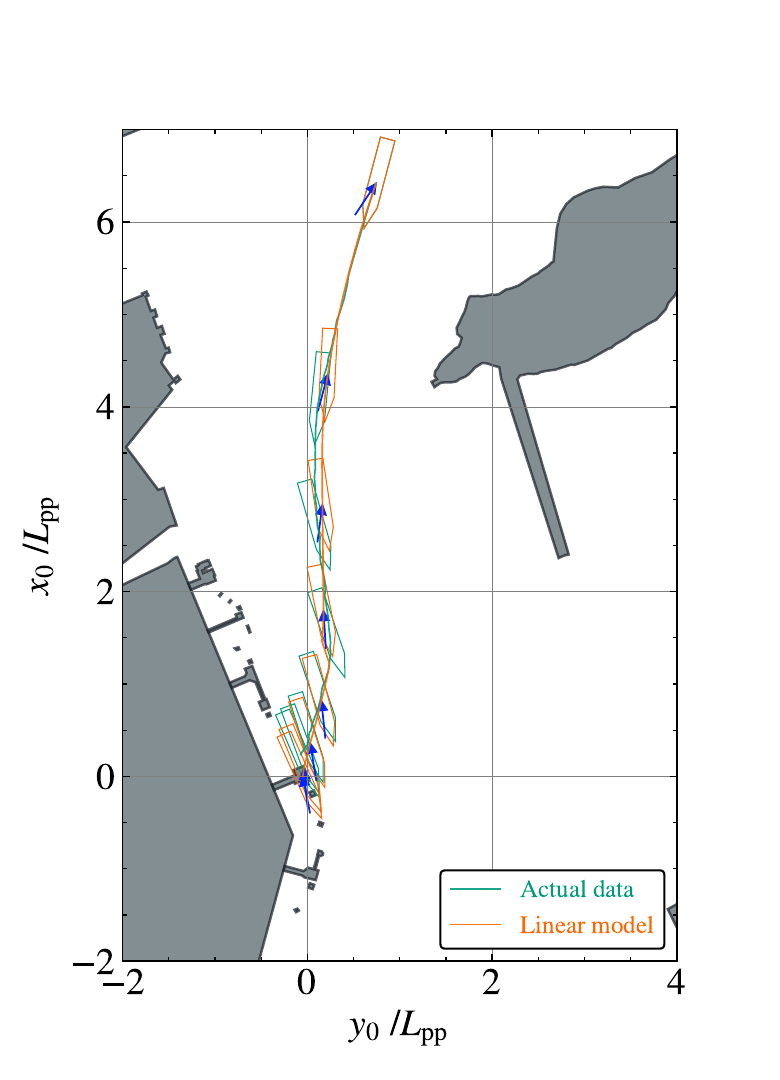}
    \label{fig: port1 traj}
\end{minipage}
\begin{minipage}[h!]
    {\linewidth}
        \centering
    \includegraphics[keepaspectratio, width = 0.95\hsize]{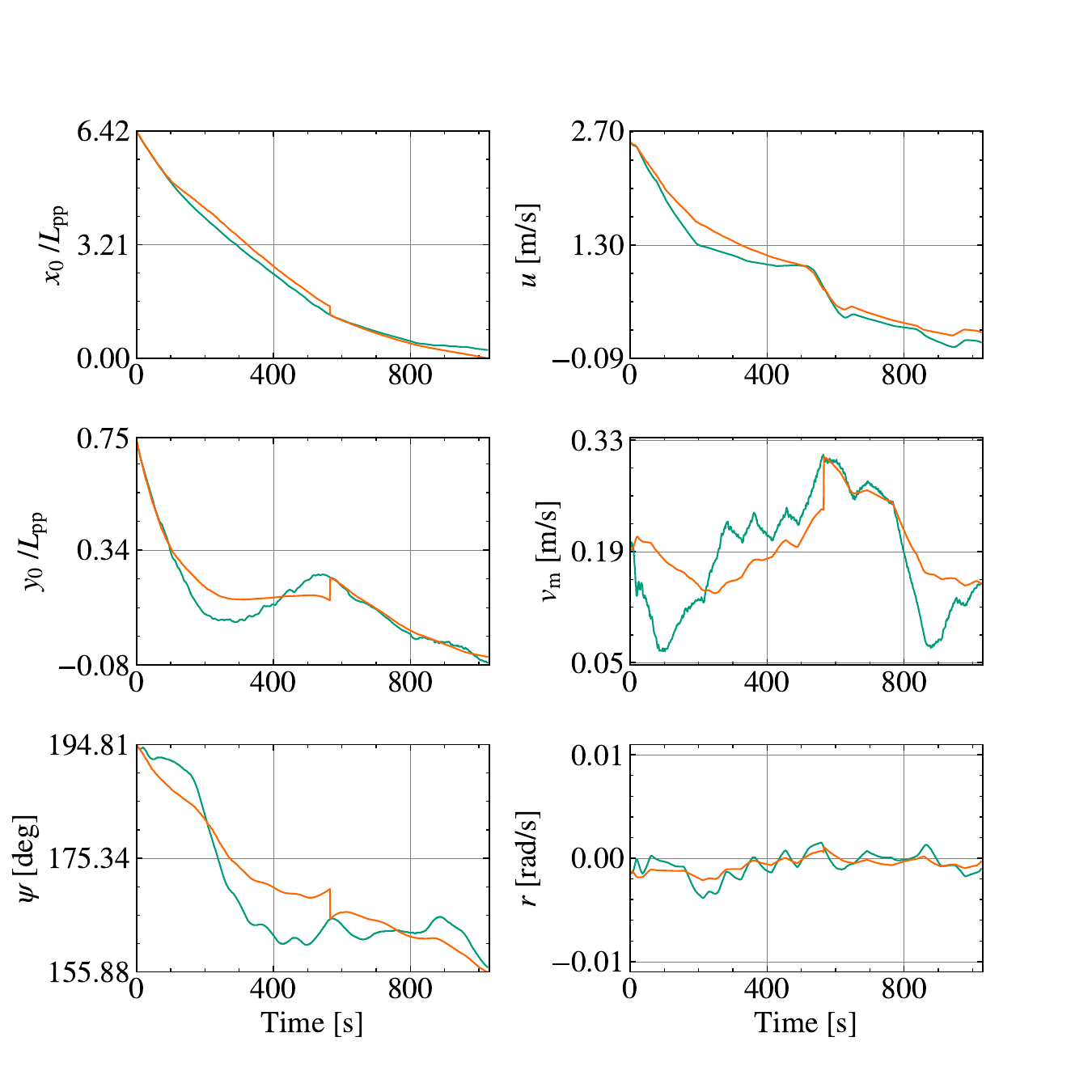}
    \label{fig: port1 states}
\end{minipage}
\caption{Port 1 - Comparison between the full-scale ship data (teal) and the model-simulated (orange) and the corresponding time-series state variables. The blue arrow denotes the relative wind direction, $\UA$.
The model-predicted trajectories exhibit minor deviations from the measured ship states; however, following re-initialization at 500s, the simulated trajectories become nearly identical to those of the actual ship.}
\label{fig: port1 traj and states}
\end{figure}

\begin{figure}[htbp]
\begin{minipage}[h!]
    {\linewidth}
        \centering
    \includegraphics[keepaspectratio, width = 0.95\hsize]{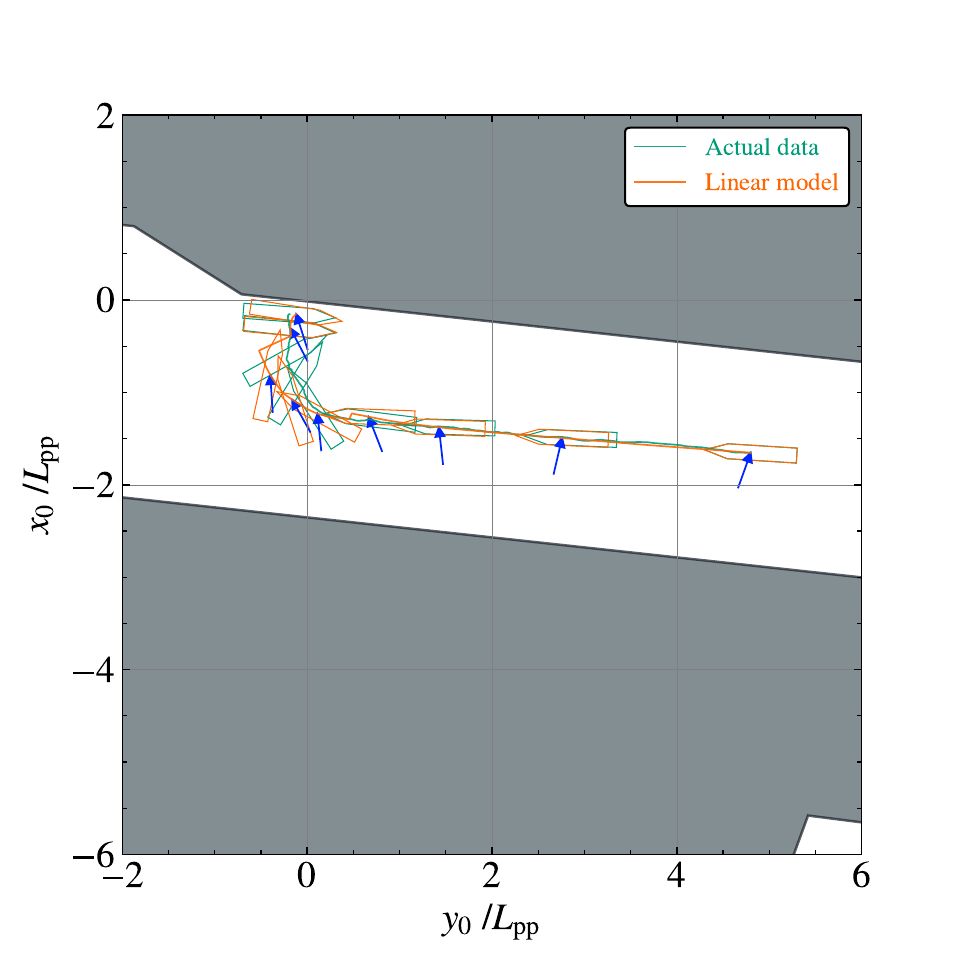}
    \label{fig: port2 traj}
\end{minipage}
\begin{minipage}[h!]
    {\linewidth}
        \centering
    \includegraphics[keepaspectratio, width = 0.95\hsize]{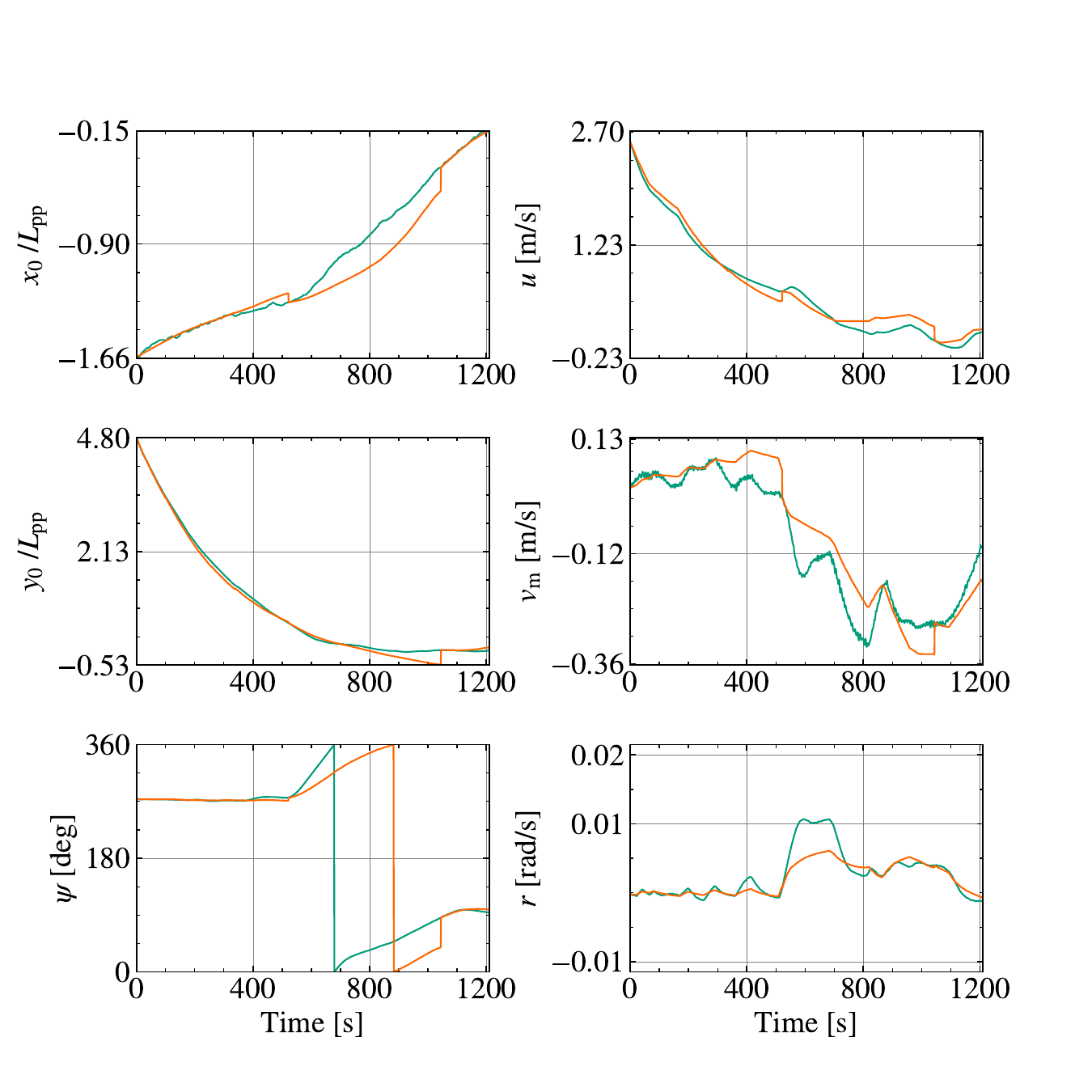}
    \label{fig: port2 states}
\end{minipage}
\caption{Port 2 - Comparison between the full-scale ship data (teal) and the model-simulated (orange) and the corresponding time-series state variables. The blue arrow denotes the relative wind direction, $\UA$. The deviation between simulated and measured trajectories at the initial 500s initialization is barely discernible. Although a more pronounced deviation is observed at the second re-initialization (1000s), the model-generated trajectories become identical to those of the actual ship immediately following the re-initialization.}
\label{fig: port2 traj and states}
\end{figure}

\begin{figure}[htbp]
\begin{minipage}[h!]
    {\linewidth}
        \centering
    \includegraphics[keepaspectratio, width = 0.95\hsize]{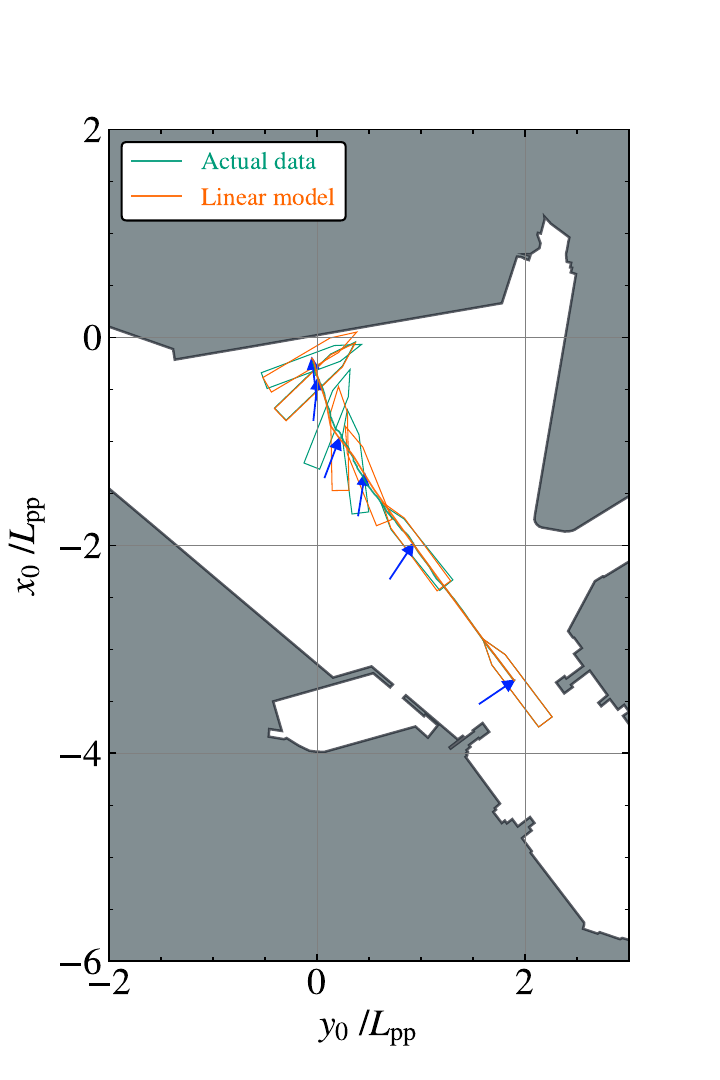}
    \label{fig: port3 traj}
\end{minipage}
\begin{minipage}[h!]
    {\linewidth}
        \centering
    \includegraphics[keepaspectratio, width = 0.95\hsize]{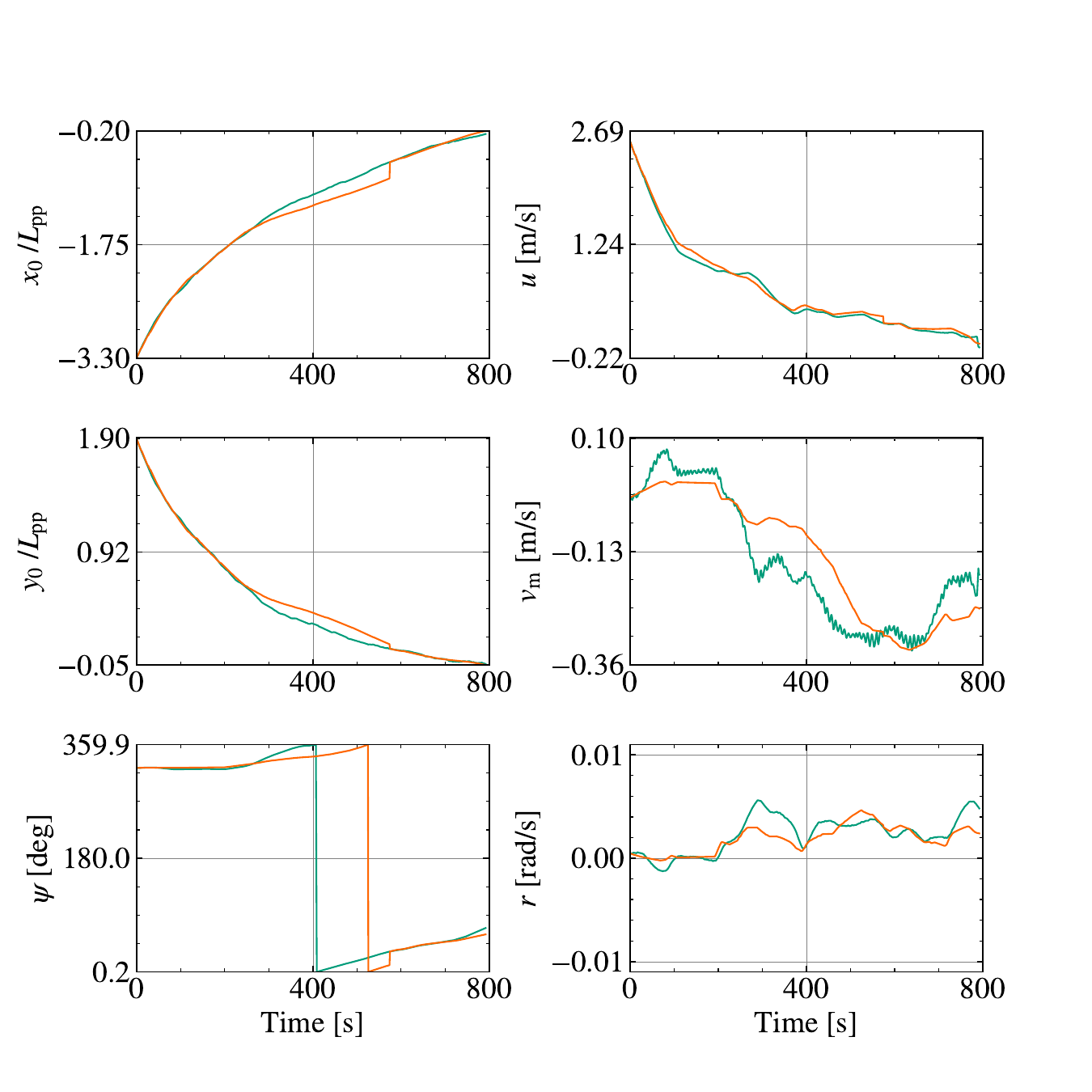}
    \label{fig: port3 states}
\end{minipage}
\caption{Port 3 - Comparison between the full-scale ship data (teal) and the model-simulated (orange) and the corresponding time-series state variables. The blue arrow denotes the relative wind direction, $\UA$. The trajectory is approximately 800s long and at the initial re-initialization (500s), the deviation of the model-predicted trajectories from the measured ship states is barely discernible. Thereafter, the simulated trajectories become nearly identical to those of the actual ship.}
\label{fig: port3 traj and states}
\end{figure}

\begin{figure}[htbp]
\begin{minipage}[h!]
    {\linewidth}
        \centering
    \includegraphics[keepaspectratio, width = 0.95\hsize]{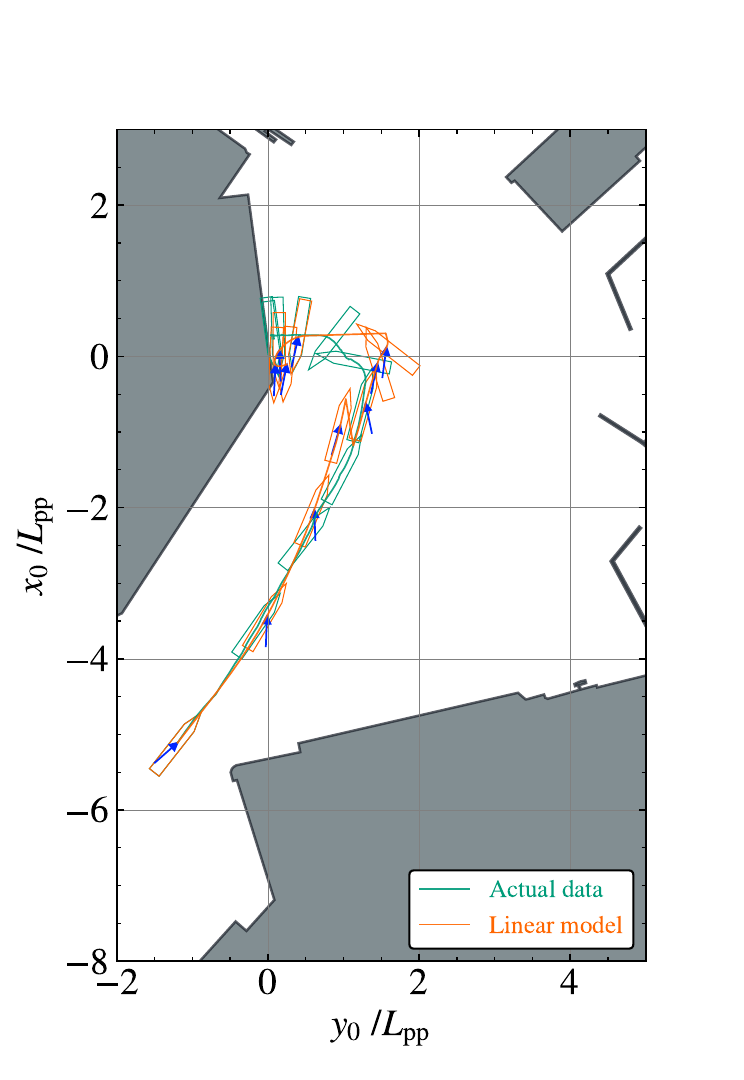}
    \label{fig: port4 traj}
\end{minipage}
\begin{minipage}[h!]
    {\linewidth}
        \centering
    \includegraphics[keepaspectratio, width = 0.95\hsize]{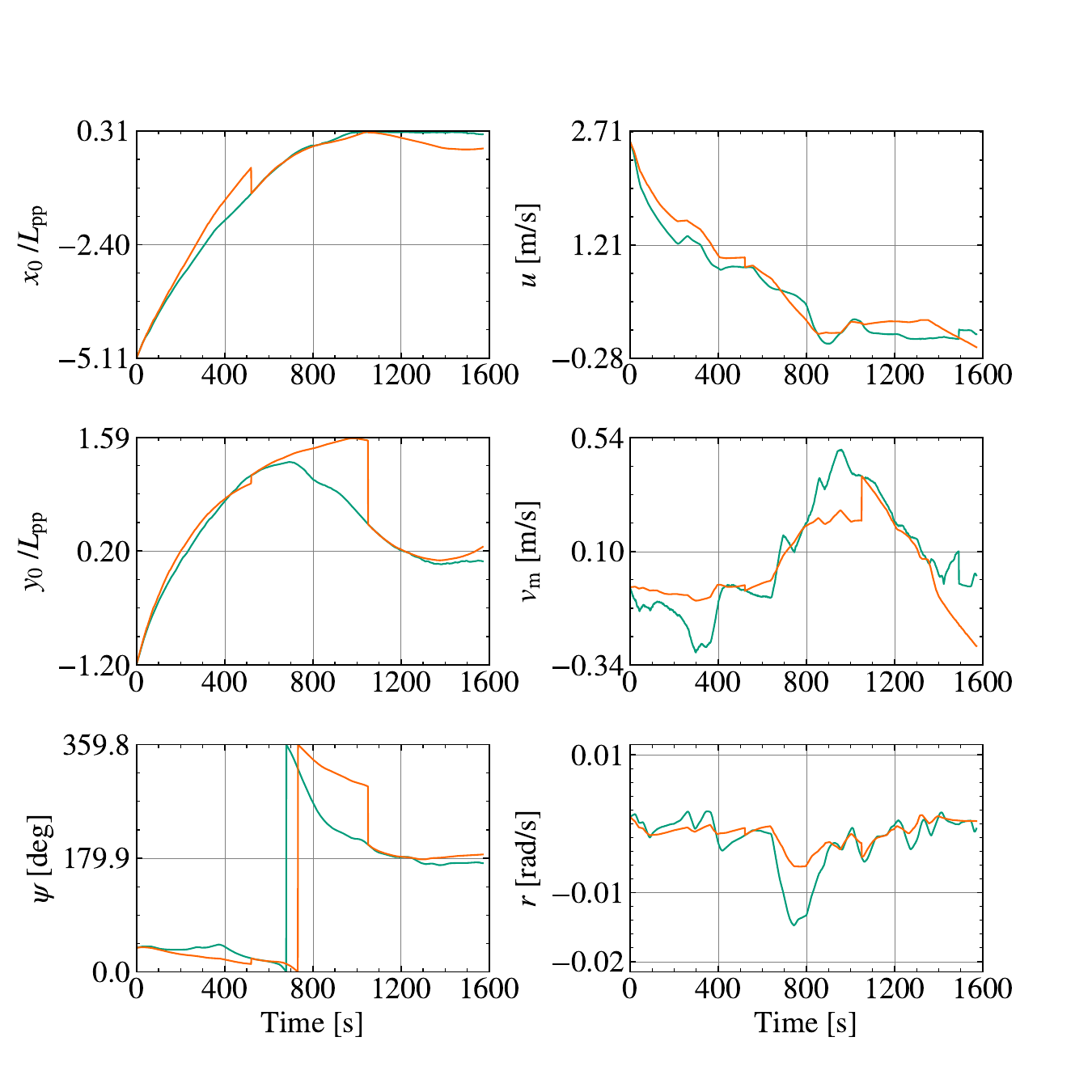}
    \label{fig: port4 states}
\end{minipage}
\caption{Port 4 - Comparison between the full-scale ship data (teal) and the model-simulated (orange) and the corresponding time-series state variables. The blue arrow denotes the relative wind direction, $\UA$. Among the five test cases, the trajectory at Port 4 is the longest, extending to approximately 1600s. Longitudinal deviations are most pronounced following the first re-initialization at 500s, whereas lateral deviations become more evident after the second re-initialization at 1000s. After the second re-initialization, the model-generated trajectories converge to those of the actual ship.}
\label{fig: port4 traj and states}
\end{figure}

\begin{figure}[htbp]
\begin{minipage}[h!]
    {\linewidth}
        \centering
    \includegraphics[keepaspectratio, width = 0.95\hsize]{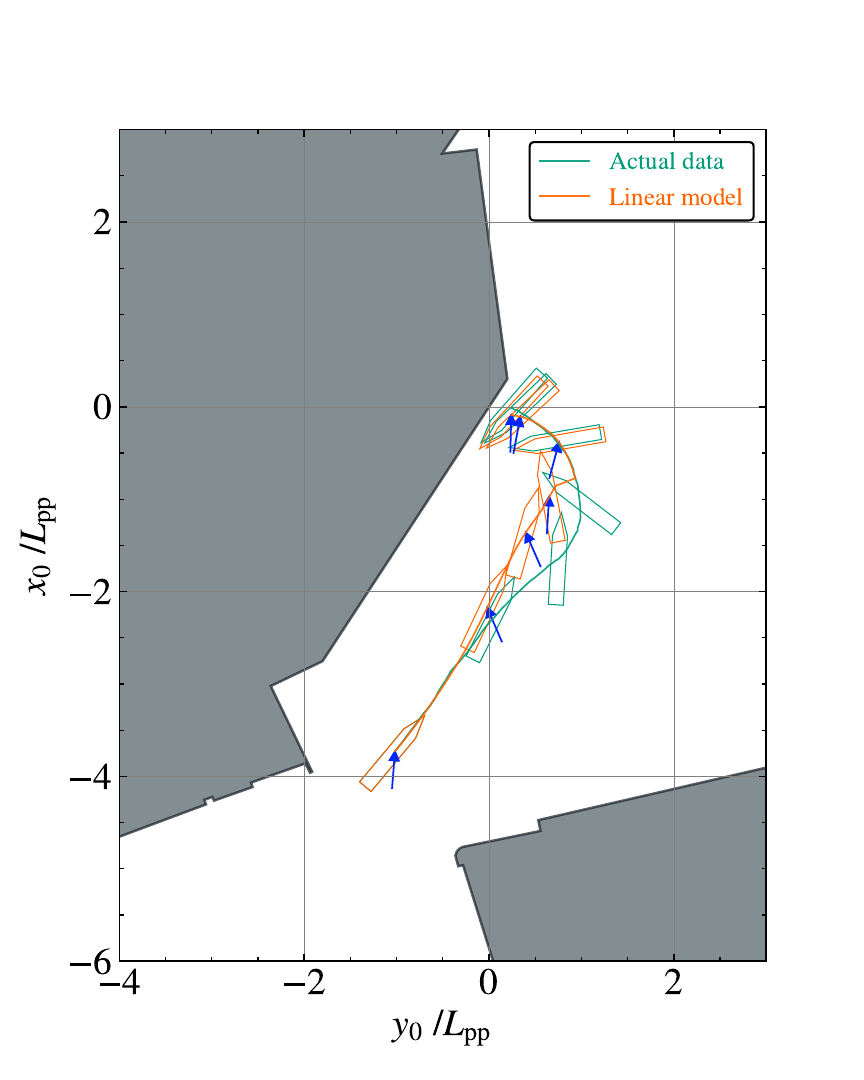}
    \label{fig: port5 traj}
\end{minipage}
\begin{minipage}[h!]
    {\linewidth}
        \centering
    \includegraphics[keepaspectratio, width = 0.95\hsize]{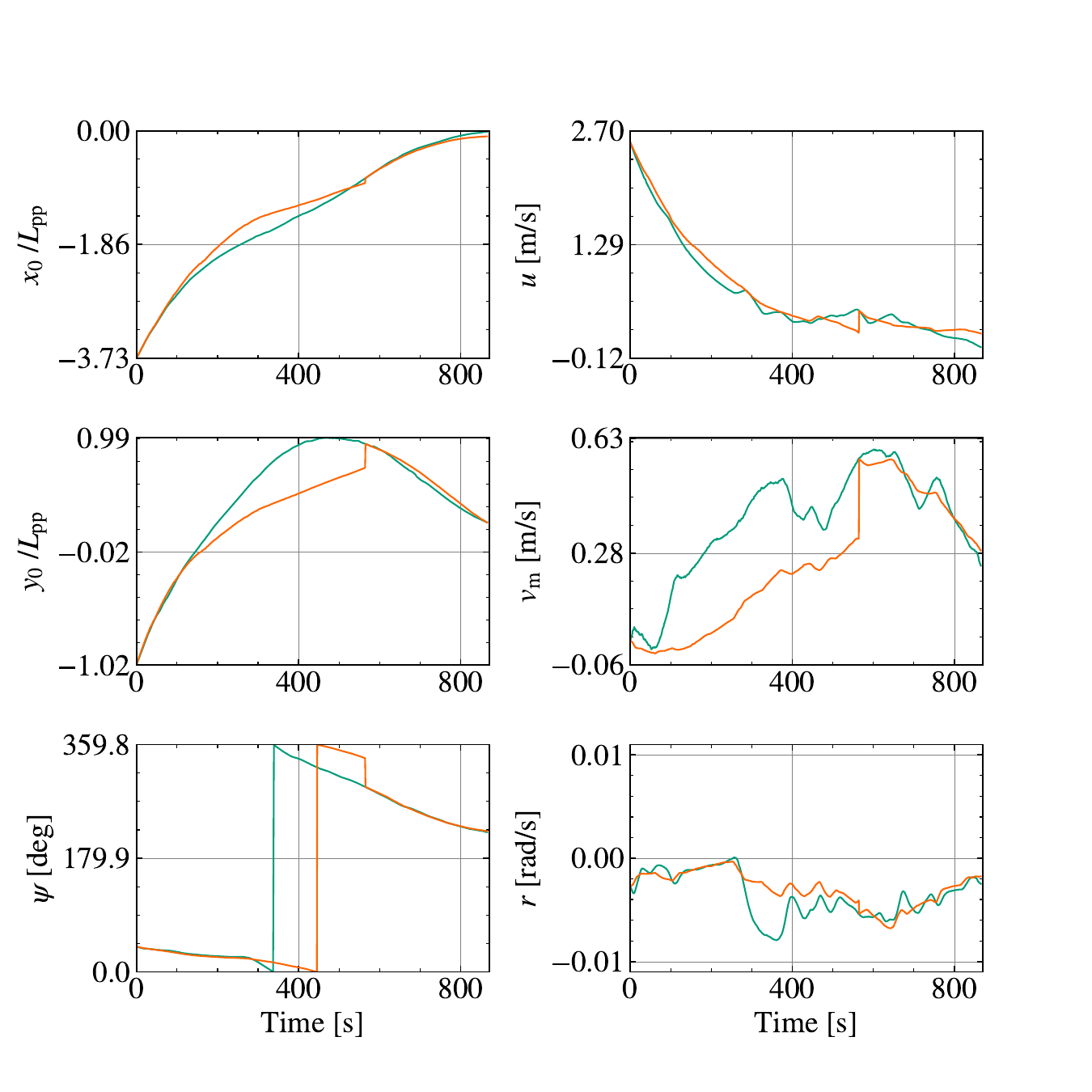}
    \label{fig: port5 states}
\end{minipage}
\caption{Port 5 - Comparison between the full-scale ship data (teal) and the model-simulated (orange) and the corresponding time-series state variables. The blue arrow denotes the relative wind direction, $\UA$. The trajectory duration at Port 5 is comparable to that of Port 3. However, in contrast to Port 3, the lateral and pose deviations at the initial re-initialization (500s) are considerably more pronounced. This could be attributed to the difference in berthing pattern and actuator arrangement in preparation for berthing. As with the other test cases, the model-generated trajectories converge to the measured ship states after re-initialization.}
\label{fig: port5 traj and states}
\end{figure}

Furthermore, model validation was performed using the $\mathcal{D}_{\mathrm{test}}$ data set, with the comparative trajectories for each port presented in \cref{fig: port1 traj and states} through \cref{fig: port5 traj and states}. The results demonstrate a close agreement between the simulated and the actual ship trajectories, confirming the model's capability to replicate low-speed ship dynamics. 

\section{Discussion and Limitations}\label{sec: discussion}
Simulation results across multiple berthing scenarios, presented in \cref{fig: port1 traj and states} through \cref{fig: port5 traj and states}, confirm the predictive capability of the proposed model. The comparative time-series plots demonstrate excellent agreement with full-scale measurements across a diverse set of berthing maneuvers. In each case, the simulation was re-initialized with true ship states at 500s intervals, with the total number of re-initializations dictated by the varying trajectory lengths across ports. Notably, except for Port 4, the deviations between the model-generated and the actual ship trajectories at re-initialization points are almost negligible. Moreover, across all test cases, the model converges to the actual ship trajectories following the final re-initialization, excellently replicating the final phase of each berthing operation. This convergence is particularly significant, as it validates the model’s capacity to capture low-speed maneuvering dynamics.

\begin{figure*}[htbp]
    \centering
    \includegraphics[width = 1.0\textwidth]{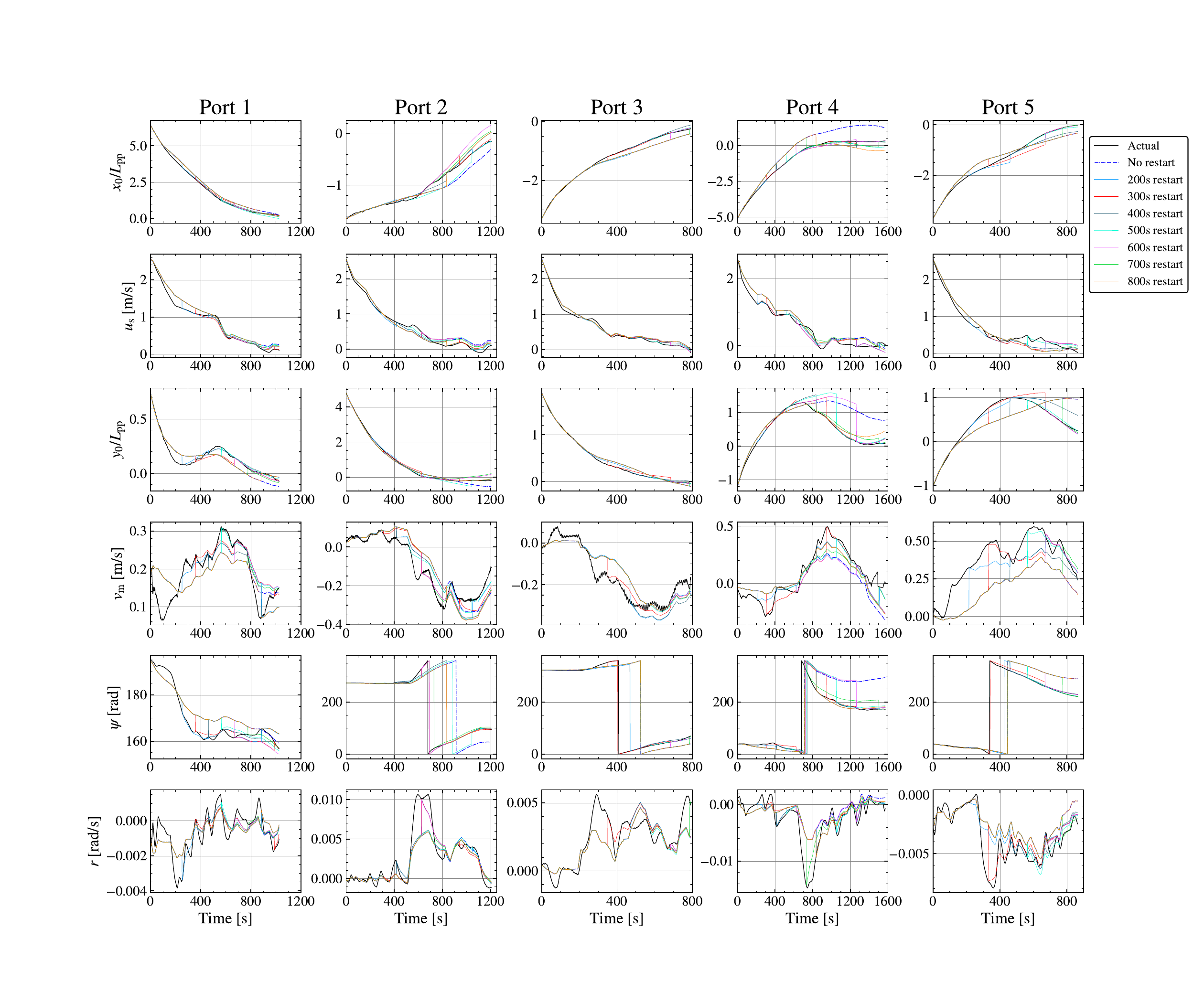}
    \caption{Comparison of model-predicted trajectories under varying re-initialization intervals against actual ship trajectories (solid black) and open-loop model-simulated trajectories without state reset (dashed blue).}
    \label{fig: restart comparison}
\end{figure*}

Further, to assess the model’s sensitivity to re-initialization frequency, simulations were conducted using restart intervals of 200, 300, 400, 500, 600, 700, and 800s. Notably, for berthing maneuvers lasting approximately 800 seconds, the 800s restart effectively corresponds to a simulation without re-initialization. As depicted in \cref{fig: restart comparison}, increased deviations between model predictions and measured ship data become apparent in time-series exceeding 1000 seconds, particularly in the lateral velocity $\vm$, yaw angle $\psi$, and yaw rate $r$. This behavior is expected, as lateral and turning dynamics are inherently nonlinear and remain among the most challenging to model accurately. 

Moreover, the parameter identification process for the proposed linear model exhibits considerable computational efficiency. As illustrated in \cref{fig: obj func and param limits}, the objective function converges rapidly during optimization, which stands in contrast to the identification of complex nonlinear models, which can require extensive computation time to converge to a global minimum. Although the linear parameter space still contains local minima, the model's simpler structure allows it to converge quickly to a global minimum. This offers a clear practical advantage in terms of computational cost and practicality for model identification and potential online adaptation.

In this study, the use of full-scale ship navigation data eliminated scaling effects and associated uncertainties inherent in model-test-based identification approaches, yielding parameter estimates that directly reflect true ship dynamics without the need for extrapolation. Furthermore, as discussed in \cref{sec: data curation}, this study utilized training data that was carefully curated to mitigate multicollinearity among key state variables such as yawrate ($r$) and sway velocity ($\vm$). This precaution prevents parameter cancellation and preserves the identifiability of individual model parameters, particularly common with polynomial and hydrodynamic maneuvering models \cite{Hwang1982, measures_luo_2017}. 

A primary limitation in this study is that the model's formulation and identified parameters are intrinsically linked to the specific actuation system of the subject ship, which is equipped with a vectwin rudder system. As such, the results of this study are not directly applicable to conventional ships with standard rudder–propeller configurations or to other overactuated ships. Consequently, the generalized application of the proposed model would necessitate modification to the model and re-identification of parameters tailored to the actuation characteristics of the target ship.

\section{Conclusion}\label{sec: conclusion}
This study presented a linear maneuvering model specifically formulated for low-speed maneuvering. Validation against full-scale ship data confirms that the proposed approach captures the low-speed ship dynamics with notable computational efficiency, while avoiding the scaling effects inherent in conventional model-test methodologies. The findings demonstrate that a structurally simplified formulation can serve as a viable and practical alternative to complex nonlinear models, offering a dynamically consistent and computationally tractable representation of near-zero-speed vessel behavior. Finally, the model directly supports downstream  application such as trajectory planning, robust control synthesis, and predictive collision avoidance for autonomous berthing and unberthing applications.
  
\begin{acknowledgements}
\normalsize{This research was conducted in collaboration with Japan Hamworthy Co. The authors would like to express their deepest gratitude to Japan Hamworthy Co. This work was also supported by a Grant-in-Aid for Scientific Research from the Japan Society for Promotion of Science (JSPS KAKENHI Grant Number 22H01701).}
\end{acknowledgements}

%
\section*{Conflict of interest}
The authors declare that they have no conflict of interest.

\section*{Data Availability Statement}
Due to confidentiality agreements, the authors are unable to make the data used in this study publicly available.

\bibliographystyle{spphys}       
\bibliography{main.bib}   

@article{Miyauchi2022SI,
   author = {Yoshiki Miyauchi and Atsuo Maki and Naoya Umeda and Dimas M. Rachman and Youhei Akimoto},
   doi = {10.1007/s00773-022-00889-3},
   issn = {0948-4280},
   journal = {Journal of Marine Science and Technology},
   month = {6},
   title = {System parameter exploration of ship maneuvering model for automatic docking/berthing using CMA-ES},
   url = {http://arxiv.org/abs/2111.06124 https://link.springer.com/article/10.1007/s00773-022-00889-3 https://link.springer.com/10.1007/s00773-022-00889-3},
   year = {2022}
}

@article{KOBAYASHI1994,
   author = {Hiroaki KOBAYASHI and Atsushi ISHIBASHI and Kenji SHIMOKAWA and Yuichi SHIMURA},
   doi = {10.9749/jin.91.263},
   issn = {0388-7405},
   issue = {0},
   journal = {The Journal of Japan Institute of Navigation},
   pages = {263-270},
   title = {A Study on Mathematical Model for the Maneuvering Motions of Twin-propeller Twin-rudder Ship : In Reference to the Maneuvering Motion from Ordinary Speed Range to Low Speed Range},
   volume = {91},
   url = {https://www.jstage.jst.go.jp/article/jin/91/0/91_KJ00004721687/_article/-char/ja/},
   year = {1994}
}

@article{YOSHIMURA1989,
   author = {Yasuo YOSHIMURA and Hitoshi SAKURAI},
   doi = {10.14856/kansaiks.211.0_115},
   journal = {Journal of the Kansai Society of Naval Architects, Japan},
   pages = {115-126},
   title = {Mathematical Model for the Manoeuvring Ship Motion in Shallow Water (3rd Report) : Manoeuvrability of a Twin-propeller Twin rudder ship},
   volume = {211},
   year = {1989}
}

@article{Lee1988,
   author = {Seung Keon Lee and Masataka Fujino and Toichi Fukasawa},
   doi = {10.2534/jjasnaoe1968.1988.109},
   issn = {0514-8499},
   issue = {163},
   journal = {Journal of the Society of Naval Architects of Japan},
   pages = {109-118},
   title = {A Study on the Manoeuvring Mathematical Model for a Twin-Propeller Twin-Rudder Ship},
   volume = {1988},
   url = {http://joi.jlc.jst.go.jp/JST.Journalarchive/jjasnaoe1968/1988.109?from=CrossRef},
   year = {1988}
}

@article{Kose1979a,
   author = {Kuniji Kose and Toshiro Saeki},
   doi = {10.2534/jjasnaoe1968.1979.146_229},
   issn = {1884-2070},
   issue = {146},
   journal = {Journal of the Society of Naval Architects of Japan},
   pages = {229-236},
   title = {On a New Mathematical Model of Manoeuvring Motions of a Ship},
   volume = {1979},
   year = {1979}
}

@article{Ogawa1978b,
   author = {A Ogawa and H Kasai},
   doi = {10.3233/ISP-1978-2529202},
   issn = {1566-2829},
   issue = {292},
   journal = {International Shipbuilding Progress},
   pages = {306-319},
   publisher = {IOS Press},
   title = {On the mathematical model of manoeuvring motion of ships},
   volume = {25},
   year = {1978}
}

@article{YOSHIMURA1988,
   author = {Yasuo YOSHIMURA and Yasuo},
   doi = {10.14856/kansaiks.210.0_77},
   issn = {0389-9101},
   journal = {Journal of the Kansai Society of Naval Architects, Japan},
   pages = {77-84},
   publisher = {The Japan Society of Naval Architects and Ocean Engineers},
   title = {Mathematical Model for the Manoeuvring Ship Motion in Shallow Water : 2nd Report : Mathematical Model at Slow Forward Speed},
   volume = {210},
   url = {http://ci.nii.ac.jp/naid/110003873057/en/},
   year = {1988}
}

@article{HIRANO1985,
   author = {Masayoshi HIRANO and Junshi TAKASHINA and Shuko MORIYA and Yoshiaki NAKAMURA},
   doi = {10.14856/wjsna.69.0_101},
   journal = {TRANSACTIONS OF THE WEST-JAPAN SOCIETY OF NAVAL ARCHITECTS},
   pages = {101-110},
   title = {An Experimental Study on Maneuvering Hydrodynamic Forces in Shallow Water},
   volume = {69},
   year = {1985}
}

@article{Ueno2017,
   author = {Michio Ueno and Ryosuke Suzuki and Yoshiaki Tsukada},
   doi = {10.1016/j.oceaneng.2016.12.001},
   issn = {00298018},
   journal = {Ocean Engineering},
   month = {1},
   pages = {260-273},
   publisher = {Elsevier Ltd},
   title = {Estimation of stopping ability of full-scale ship using free-running model},
   volume = {130},
   year = {2017}
}

@article{Ueno2015a,
   author = {Michio Ueno and Yoshiaki Tsukada},
   doi = {10.1016/J.OCEANENG.2015.09.041},
   issn = {0029-8018},
   journal = {Ocean Engineering},
   pages = {495-506},
   publisher = {Pergamon},
   title = {Rudder effectiveness and speed correction for scale model ship testing},
   volume = {109},
   year = {2015}
}

@article{Ueno2014a,
   author = {Michio Ueno and Yoshiaki Tsukada and Yasushi Kitagawa},
   doi = {10.1016/j.oceaneng.2014.10.006},
   issn = {00298018},
   journal = {Ocean Engineering},
   pages = {267-284},
   publisher = {Elsevier},
   title = {Rudder effectiveness correction for scale model ship testing},
   volume = {92},
   year = {2014}
}

@article{Ueno2001,
   author = {Michio Ueno and Tadashi Nimura and Hideki Miyazaki and Toshifumi Fujiwara and Koji Nonaka and Hideo Yabuki},
   doi = {10.2534/jjasnaoe1968.2001.71},
   issn = {1884-2070},
   issue = {189},
   journal = {Journal of the Society of Naval Architects of Japan},
   pages = {71-80},
   title = {Model Experiment and Sea Trial for Investigating Manoeuvrability of a Training Ship},
   volume = {2001},
   url = {http://joi.jlc.jst.go.jp/JST.Journalarchive/jjasnaoe1968/2001.71?from=CrossRef},
   year = {2001}
}

@inproceedings{Crane1979,
   author = {C L Crane},
   booktitle = {SNAME Transactions},
   pages = {251-283},
   title = {MANEUVERING TRIALS OF THE 278,000 DWT ESSO OSAKA IN SHALLOW AND DEEP WATERS},
   volume = {87},
   year = {1979}
}

@article{FUJINO1978,
   author = {Masataka FUJINO and Atsushi KIRITA},
   issn = {0389-9101},
   journal = {Journal of the Kansai Society of Naval Architects, Japan},
   pages = {57-70},
   publisher = {The Japan Society of Naval Architects and Ocean Engineers},
   title = {On the Manoeuvrability of Ships while Stopping by Adverse Rotation of Propeller : 1st Report},
   volume = {169},
   url = {http://ci.nii.ac.jp/naid/110003873312/en/},
   year = {1978}
}

@article{Yoshimura1978,
   author = {Yasuo Yoshimura and Kensaku Nomoto},
   doi = {10.2534/jjasnaoe1968.1978.144_57},
   isbn = {9781787284395},
   issn = {1884-2070},
   issue = {144},
   journal = {Journal of the Society of Naval Architects of Japan},
   pages = {57-69},
   title = {Modeling of Manoeuvring Behaviour of Ships with a Propeller Idling, Boosting and Reversing},
   volume = {1978},
   url = {http://www.jstage.jst.go.jp/article/jjasnaoe1968/1978/144/1978_144_57/_article},
   year = {1978}
}

@article{Sakamoto2019,
   author = {Nobuaki Sakamoto and Kunihide Ohashi and Motoki Araki and Ken ichi Kume and Hiroshi Kobayashi},
   doi = {10.1016/j.oceaneng.2019.106257},
   issn = {00298018},
   issue = {August},
   journal = {Ocean Engineering},
   pages = {106257},
   publisher = {Elsevier Ltd},
   title = {Identification of KVLCC2 manoeuvring parameters for a modular-type mathematical model by RaNS method with an overset approach},
   volume = {188},
   year = {2019}
}

@article{Liu2018,
   author = {Yi Liu and Lu Zou and Zaojian Zou and Haipeng Guoa},
   doi = {10.1080/19942060.2018.1439773},
   issn = {1997003X},
   issue = {1},
   journal = {Engineering Applications of Computational Fluid Mechanics},
   pages = {334-353},
   title = {Predictions of ship maneuverability based on virtual captive model tests},
   volume = {12},
   url = {https://doi.org/10.1080/19942060.2018.1439773},
   year = {2018}
}

@article{Hwang1982,
   author = {Wei-Yuan Hwang},
   doi = {10.3233/ISP-1982-2933201},
   issn = {15662829},
   issue = {332},
   journal = {International Shipbuilding Progress},
   month = {4},
   pages = {90-102},
   title = {Cancellation effect and parameter identifiability of ship steering dynamics},
   volume = {29},
   url = {https://www.medra.org/servlet/aliasResolver?alias=iospress&doi=10.3233/ISP-1982-2933201},
   year = {1982}
}

@phdthesis{Hwang1980,
   author = {Wei-yuan Hwang},
   title = {Application of system identification to ship maneuvering},
   year = {1980}
}

@inproceedings{Abkowitz1980,
   author = {Martin A Abkowitz},
   booktitle = {Transactions of Society of Naval Architects and Marine Engineers 88},
   pages = {283–318},
   title = {Measurement of hydrodynamic characteristics from ship maneuvering trials by system identification},
   year = {1980}
}

@techReport{Strom-Tejsen1965a,
   author = {Jorgen Strom-Tejsen},
   title = {A digital computer technique for prediction of standard maneuvers of surface ships},
   year = {1965}
}

@techReport{Abkowitz1964a,
   author = {Martin A Abkowitz},
   title = {Lectures on ship hydrodynamics--Steering and manoeuvrability},
   year = {1964}
}

@article{unified_yoshimura_2009,
	title = {Unified Mathematical Model for Ocean and Harbour Manoeuvring},
	author = {Yoshimura, Yasuo and Nakao, Ikao and Ishibashi, Atsushi},
	year = {2009},
	researchRabbitId = {9b5ac8cc-9089-48b3-b40c-2f4987975a85}
}

@article{hydrodynamic_inoue_1981,
	title = {Hydrodynamic derivatives on ship manoeuvring},
	doi = {10.3233/ISP-1981-2832103},
	author = {Inoue, Shota and Hirano, M and Kijima, K},
	year = {1981},
	researchRabbitId = {f0aa825c-861e-4e36-b6d9-8c9625a2f872}
}

@article{measures_luo_2017,
	title = {Measures to diminish the parameter drift in the modeling of ship manoeuvring using system identification},
	doi = {10.1016/J.APOR.2017.06.008},
	author = {Luo, Weilin and Li, Xinyu},
	journal = {Applied Ocean Research},
	year = {2017},
	researchRabbitId = {5a250671-13d9-4658-827e-1b3d27add351}
}

@article{handbook_fossen_2011,
	title = {Handbook of Marine Craft Hydrodynamics and Motion Control},
	doi = {10.1002/9781119575016},
	author = {Fossen, Thor I.},
	year = {2011},
	researchRabbitId = {4afefcb0-5155-45a3-9411-a43edac0ea33}
}

@article{aspects_dubbioso_2012,
	title = {Aspects of twin screw ships semi-empirical maneuvering models},
	doi = {10.1016/J.OCEANENG.2012.03.007},
	author = {Dubbioso, Giulio and Viviani, Michele},
	journal = {Ocean Engineering},
	year = {2012},
	researchRabbitId = {d48cc570-3f21-41c4-b6f7-2bf6f477f1da}
}

@article{data_deogaonkar_2023,
	title = {Data Driven Identification of Ship Maneuvering Coefficients},
	doi = {10.1115/OMAE2023-104644},
	author = {Deogaonkar, Vallabh and Jadhav, Aditya Kailas and Ramachandran, Krishnavelu and Somayajula, Abhilash Sharma},
	journal = {Ocean Engineering},
	year = {2023},
	researchRabbitId = {101cccfa-4c0e-4586-99e8-6c8896cd2e9a}
}

@article{discovering_hasan_2025,
	title = {Discovering ship maneuvering models from data},
	doi = {10.1007/S00773-024-01045-9},
	author = {Hasan, Agus},
	journal = {Journal of Marine Science and Technology},
	year = {2025},
	researchRabbitId = {1745d164-c032-4f32-9b39-8d416842c5d2}
}

@article{simplified_biancardi_1988,
	title = {A simplified mathematical model for an onboard maneuvering simulator},
	doi = {10.1177/003754978805100207},
	author = {Biancardi, C G},
	journal = {Simulation},
	year = {1988},
	researchRabbitId = {fcb89ec0-22a4-463c-a910-40b36b2afd63}
}

@article{measurement_abkowitz_1980,
	title = {MEASUREMENT OF HYDRODYNAMIC CHARACTERISTICS FROM SHIP MANEUVERING TRIALS BY SYSTEM IDENTIFICATION},
	author = {Abkowitz, Martin A.},
	year = {1980},
	researchRabbitId = {b1c3a9ec-52b4-4365-b235-44332d04f29e}
}

@article{mathematical_kose_1984,
	title = {On a Mathematical Model of Maneuvering Motions of Ships in Low Speeds},
	doi = {10.2534/JJASNAOE1968.1984.132},
	author = {Kose, Kuniji and HINATA, Hiroyoshi and Hashizume, Yasuhisa and Futagawa, Eijiro},
	year = {1984},
	researchRabbitId = {32e7d163-03e0-4672-8362-6223cd4125c3}
}

@article{introduction_yasukawa_2015,
	title = {Introduction of MMG standard method for ship maneuvering predictions},
	doi = {10.1007/S00773-014-0293-Y},
	author = {Yasukawa, Hironori and Yoshimura, Yasuo},
	journal = {Journal of Marine Science and Technology},
	year = {2015},
	researchRabbitId = {c945777f-d75f-4864-8d51-e785f52a2348}
}

@article{Mwange2025,
   author = {Agnes N. Mwange and Yoshiki Miyauchi and Taichi Kambara and Hiroaki Koike and Kazuyoshi Hosogaya and Atsushi Ishibashi and Atsuo Maki},
   doi = {10.1007/s00773-025-01098-4},
   issn = {09484280},
   journal = {Journal of Marine Science and Technology},
   publisher = {Springer},
   title = {Quantitative evaluation of full-scale ship maneuvering characteristics during berthing and unberthing},
   year = {2025}
}

@article{rachman2023experimental,
  title={Experimental low-speed positioning system with VecTwin rudder for automatic docking (berthing)},
  author={Rachman, Dimas M and Aoki, Yusuke and Miyauchi, Yoshiki and Umeda, Naoya and Maki, Atsuo},
  journal={Journal of Marine Science and Technology},
  volume={28},
  number={3},
  pages={689--703},
  year={2023},
  publisher={Springer}
}

@inproceedings{sakamoto2017modified,
  title={Modified box constraint handling for the covariance matrix adaptation evolution strategy},
  author={Sakamoto, Naoki and Akimoto, Youhei},
  booktitle={Proceedings of the Genetic and Evolutionary Computation Conference Companion},
  pages={183--184},
  year={2017}
}

@article{maki2020application,
  title={Application of optimal control theory based on the evolution strategy (CMA-ES) to automatic berthing},
  author={Maki, Atsuo and Sakamoto, Naoki and Akimoto, Youhei and Nishikawa, Hiroyuki and Umeda, Naoya},
  journal={Journal of Marine Science and Technology},
  volume={25},
  pages={221--233},
  year={2020},
  publisher={Springer}
}
\end{document}